\documentclass[iop,twocolappendix,numberedappendix]{emulateapj}

\usepackage{amsmath}
\usepackage{txfonts}
\usepackage{graphicx}
\usepackage{hyperref}
\usepackage[utf8]{inputenc}

\newcommand{\td}[2]{\frac{d#1}{d#2}}
\newcommand{\eps}{\epsilon}
\newcommand{\e}[1]{\times10^{#1}}

\begin{document}

\title{Exploring the Properties of Choked Gamma-Ray Bursts with IceCube's High Energy Neutrinos}
\shorttitle{Neutrinos From Choked Gamma-Ray Bursts}

\shortauthors{Denton and Tamborra}

\author{Peter B.~Denton\altaffilmark{1} and
Irene Tamborra\altaffilmark{1,2}}

\affil{\altaffilmark{1} Niels Bohr International Academy, Niels Bohr Institute, University of Copenhagen, Blegdamsvej 17, 2100, Copenhagen, Denmark\\
\altaffilmark{2} DARK, Niels Bohr Institute, University of Copenhagen, Juliane Maries Vej 30, 2100, Copenhagen, Denmark}


\begin{abstract}
Long duration gamma-ray bursts (GRBs) have often been considered as the natural evolution of some core-collapse supernova (CCSN) progenitors.
However, the fraction of CCSNe linked to astrophysical jets and their properties are still poorly constrained.
While any successful astrophysical jet harbored in a CCSN should produce high energy neutrinos, photons may be able to successfully escape the stellar envelope only for a fraction of progenitors, possibly leading to the existence of high-luminosity, low-luminosity and not-electromagnetically bright (``choked'') GRBs. By postulating a CCSN--GRB connection, we accurately model the jet physics within the internal-shock GRB model
and assume scaling relations for the GRB parameters that depend on the Lorentz boost factor $\Gamma$. The IceCube high energy neutrino flux is then employed as an upper limit of the neutrino background from electromagnetically bright and choked GRBs to constrain the jet and the progenitor properties. 
The current IceCube data set is compatible with up to $1\%$ of all CCSNe harboring astrophysical jets. Interestingly, those jets are predominantly choked.
Our findings suggest that neutrinos can be powerful probes of the burst physics and can provide major insights on the CCSN--GRB connection.
\end{abstract}


\section{Introduction}
Gamma ray bursts (GRBs) are among the most energetic astrophysical transients~\citep{Meszaros:2006rc,Kumar:2014upa,Meszaros:2015krr}.
GRBs are expected to be sources of high energy neutrinos produced through hadronic and lepto-hadronic interactions.
Neutrinos from GRBs could be possibly detected by neutrino telescopes.
However, targeted searches of neutrinos from high-luminosity GRBs (HL-GRBs) have reported evidence for a lack of statistically significant spatial and timing correlation of neutrino data \citep{Schmid:2015msr,Aartsen:2017wea} constraining the proposed theoretical models \citep{Baerwald:2011ee,He:2012tq,Bartos:2013hf}.

At the same time, a flux of astrophysical neutrinos has been detected by the IceCube Neutrino Observatory \citep{Aartsen:2013bka,Aartsen:2013jdh,
Aartsen:2014gkd,Aartsen:2014muf, Aartsen:2015knd, Aartsen:2015rwa,Aartsen:2015zva}.
While the flux is held to be predominantly extragalactic~\citep{Palladino:2016zoe,Denton:2017csz,Aartsen:2017ujz}, its origin is currently unknown.
In this context, ``choked'' GRBs (see e.g.~\cite{Meszaros:2001ms,Ando:2005xi}) have been considered as possible sources of some of the IceCube neutrinos~\citep{Murase:2013ffa,Tamborra:2015fzv,Tamborra:2015qza,Senno:2015tsn,Murase:2015xka,Senno:2017vtd}. 

A choked jet is one where the jet is successful in accelerating particles, but electromagnetic radiation is unsuccessful in escaping the stellar envelope. Choked GRBs may even be more abundant than the GRBs ordinarily observed in photons \citep{Meszaros:2001ms,Razzaque:2002kb,Razzaque:2003uw,Razzaque:2003uv,Ando:2005xi,Horiuchi:2007xi,Murase:2013ffa}. Neutrinos and possibly gravitational waves may be the only messengers from these sources.

There is solid evidence that GRBs and core-collapse supernovae (CCSNe) are related~\citep{Paczynski:1997yg,Hjorth:2011zx,Modjaz:2011bm,Hjorth:2013axa,Margutti:2014gha,Sobacchi:2017wcq,Lazzati:2011ay}.
This was also foreseen in the so-called collapsar model~\citep{MacFadyen:1998vz,MacFadyen:1999mk,Woosley:2006fn} and is supported by the fact that we expect a comparable amount of energy to be released in CCSNe and GRBs.
Recent work, see e.g.~\cite{Sobacchi:2017wcq}, suggests that possibly a large fraction of jets harbored in CCSNe might not be electromagnetically visible.
In this work, we take the CCSN--GRB relationship seriously and investigate a model of astrophysical jets originating from CCSNe.

We build up a model where the same physics applies in jets that produce $\gamma$-rays and those that do not when the jet is trapped within the stellar envelope.
According to our scenario, one could think of high (HL)- and low--luminosity (LL)- GRBs as sub-classes of one larger ensemble to which choked GRBs also belong \citep{Bromberg:2011fm,Nakar:2015tma}.
Early literature \citep{Murase:2006mm,Gupta:2006jm,Liu:2011cua,Liu:2012pf,Murase:2013ffa,Tamborra:2015fzv,Tamborra:2015qza} proved that the neutrino production from LL- and choked GRBs can even be larger than the one expected for ordinary GRBs.

We explore a general model where the jet properties scale as a function of the Lorentz boost factor $\Gamma$ and estimate the total diffuse neutrino background from both bright and choked jets, by linking choked GRBs with observed GRBs under the same model.
To do this, we perform detailed simulations within the internal-shock GRB model and vary the jet parameters to provide a realistic picture of the diffuse flux from the whole jet population. We include both $pp$ and $p\gamma$ interactions as well as all relevant cooling processes for protons and intermediate accelerated particles and adopt energy-dependent cross sections for all cooling processes relevant to our purpose. 

For the sake of completeness, we distinguish among two GRB models. The first one is a ``simple'' GRB model that is based on the commonly used scaling law $\theta_j=1/\Gamma$ applied to the whole jet, with $\theta_j$ being the half opening angle; in this model $\Gamma$ varies across the GRB population. The second one is an ``advanced'' GRB model which contains a $\Gamma$--dependent GRB population distribution as does the simple model, but also contains a distribution of $\Gamma$ within the jet.
Finally, we use IceCube data to define upper limits on the jet energy and the fraction of CCSNe that form jets.

The outline of the paper is as follows.
In Sec.~\ref{sec:neutrino flux}, we first review standard jet physics and explain how each jet parameter is related to the others.
We then calculate the flux as seen at Earth.
We introduce the simple GRB model in Sec.~\ref{sec:constant Gamma} and calculate the diffuse intensity normalized to the CCSN and HL-GRB rates.
The advanced GRB model is presented in Sec.~\ref{sec:variable Gamma}.
Then we compare the predicted diffuse intensities within the two GRB models to IceCube's data in Sec.~\ref{sec:constraints} and discuss our findings in Sec.~\ref{sec:discussion}. Our conclusions are reported in Sec.~\ref{sec:conclusions}.

\section{High Energy Neutrino Production in Astrophysical Jets}
\label{sec:neutrino flux}
In this Section, we will overview the properties of the astrophysical jets and discuss the relevant cooling processes affecting the protons and secondary particles.
We will then convert those jet properties into the diffuse neutrino intensity observed at the Earth.
For simplicity, we will now rely on the simple GRB model, wherein each jet is described by a single value of $\Gamma$, until otherwise specified.

\subsection{Properties of the Astrophysical Jet}
\label{ssec:jet properties}
We parameterize the astrophysical jet by the amount of kinetic energy in the jet $\tilde E_j$, bulk Lorentz factor $\Gamma$, and electron (magnetic) energy fraction $\eps_e$ ($\eps_B$). There are various internal properties that are functions of jet parameters.
We enumerate them here for reference.

First, we take the standard theoretical $\Gamma-\theta_j$ relation from special relativity, $\theta_j=1/\Gamma$ (see, e.g., \cite{Meszaros:2006rc}).
This form is often used throughout the literature with the argument that given a Lorentz boost factor $\Gamma$, the typical angular scale is $\theta_j=1/\Gamma$.
We hereby adopt a modified version of this $\Gamma-\theta_j$ relation to match observed jet angles:
\begin{equation}
\theta_j=
\begin{cases}
\frac1\Gamma&\Gamma\le100\\
\frac{30}\Gamma\qquad&\Gamma>100
\end{cases}\,.
\label{eq:thetaj}
\end{equation}
The above relation for a typical HL-GRB with $\Gamma=300$ gives an opening angle of $\theta_j=6^\circ$, consistent with observations \citep{Goldstein:2015fib}.
The break at $\Gamma=100$ is taken from \cite{Cenko:2010cg,Bregeon:2011bu,Dermer:2014vaa,Tamborra:2015qza}.
Measurements of the jet opening angle for LL-GRBs are more uncertain.
Nevertheless, for the range $\Gamma\in[3,100]$ the opening angle varies in the range $\theta_j\in[0.6^\circ,19^\circ]$ which is consistent with estimations of LL-GRBs jet opening angle reported in the literature \citep{Toma:2006iu,Liang:2006ci,Daigne:2007qz,Bromberg:2011fm,Zhang:2012jc,Nakar:2015tma}.

The magnetic field strength is given by,
\begin{equation}
\frac{B'^2}{8\pi}=4\eps_B\frac{E_j'}{V'}\,,
\end{equation}
where
the jet volume is given by,
\begin{equation}
V'=\Omega_j\tilde r_j^2c\tilde t_j\Gamma\,.
\label{eq:Volume}
\end{equation}
Note that we distinguish among three reference frames: $X$ - Earth, $\tilde X$ - star, $X'$ - jet.
Energies in each frame are related by $\tilde E=(1+z)E$, and $\tilde E=\Gamma E'$.
Times are related by $t=(1+z)\tilde t$, and $t'=\Gamma\tilde t$.
Luminosities are related by $\tilde L=(1+z)^2L$.
The solid angle for both jets is,
\begin{equation}
\Omega_j=4\pi(1-\cos\theta_j)\approx2\pi\theta_j^2\,.
\end{equation}
The internal-shock radius $\tilde r_j$ is defined as
\begin{equation}
\tilde r_j=2c\tilde t_v\Gamma^2\,.
\end{equation}
The jet variability time is taken from an empirical fit to HL-GRBs \citep{Sonbas:2014jya} and a maximum to cap the variability time for LL-GRBs,
\begin{equation}
\tilde t_v=\min(2.8\e9\Gamma^{-4.05},100)\,\text s\,,
\end{equation}
which breaks at $\Gamma=69$.
While variability time measurements of LL-GRBs are sparse, GRB060218/SN2006aj measured by Swift had a variability time of $\sim200$ s, so taking $\tilde t_v\sim100$ s as an upper limit for low $\Gamma$ jets is reasonable \citep{Campana:2006qe,Gupta:2006jm}.
The comoving photon, electron, and proton densities are,
\begin{equation}
n_\gamma'=\frac{4E_j'\eps_e}{V'E_{\gamma,{\rm b}}'}\,,\qquad
n_e'\simeq n_p'=\frac{E'_j}{V'm_pc^2}\,,
\end{equation}
where we have taken the comoving electron density as essentially the same as the comoving proton density.
The photon break energy $E'_{\gamma,{\rm b}}$ is defined in Sec.~\ref{ssec:proton and photon spectra}.

For the jet duration $\tilde t_j$, we use a power law relation to describe both longer GRBs with lower $\Gamma$'s and shorter GRBs with higher $\Gamma$'s.
Then, $\tilde t_j\propto 1/\Gamma$ normalized to $\tilde t_j=10$ s at $\Gamma=300$ \citep{Gehrels:2013xd,Lu:2017toj}.

\subsection{Conditions to Successfully Accelerate Particles in the Jet}
As was pointed out in \cite{Murase:2013ffa}, if the jet density is too high, then the velocity gain in each shock of protons will not be enough to reach canonical Fermi acceleration and the jet will never accelerate particles to high energy.
To ensure that a jet is successful in accelerating protons, we introduce the optical depth of the jet (given by the Thomson optical depth),
\begin{equation}
\tau_T'=\frac{\sigma_Tn_p'\tilde r_j}\Gamma\,,
\end{equation}
and apply the conservative constraint $\tau'_T\lesssim1$.
Jets that do not meet this constraint are considered unsuccessful and no particles are created in these jets.

Separate from the issue of whether or not a jet is successful at accelerating protons, we also determine if a jet is choked or not.
A choked jet is one in which the jet head does not escape the stellar envelope.
Under the assumption that the jet becomes collimated, we use the following definition for the jet head radius~\citep{Bromberg:2011fg},
\begin{multline}
\tilde r_h=5.4\e{10}{\rm\ cm\ }\tilde t_{j,1}^{6/5}\tilde L_{{\rm iso},52}^{2/5}\\
\times\left(\frac{\theta_j}{0.2}\right)^{-8/5}\left(\frac{M_*}{20\ M_\odot}\right)^{-2/5}R_{*,11}^{1/5}\,,
\label{eq:r_h}
\end{multline}
and estimate it for a typical Wolf-Rayet star with mass $M_*=20\ M_\odot$ and radius $R_*=R_\odot$.
The isotropic luminosity is $\tilde L_{\rm iso}=4\pi\tilde E_j/\tilde t_j\Omega_j$. Since many of the parameters in Eq.~\ref{eq:r_h} scale with $\Gamma$, we note that the overall $\Gamma$ and $\tilde E_j$ dependence of the jet head radius is $\tilde r_h \propto\tilde E_j^{2/5}\Gamma^{8/5}$.
We here use the standard notation, $Q_x=Q/10^x$ in cgs units, unless otherwise specified.
The condition for a jet to be choked in photons is when $\tilde r_h<R_*$.
If the internal shock radius is larger than the stellar radius then no cocoon can form and the jet is not collimated.
In this case, Eq.~\ref{eq:r_h} no longer applies as there is no jet head since the jet is visible (not choked).

\subsection{Particle Acceleration and Cooling Processes}
\label{ssec:cooling processes}
All of the charged particles in the jet, protons, pions, kaons, and muons lose energy to various processes.
To determine the final spectrum of neutrinos, the cooling process of each particle needs to be determined.
In different regimes of energy and $\Gamma$, as well as the other parameters, different cooling processes dominate.

\subsubsection{Protons}
\label{sssec:proton cooling}
Protons are accelerated on a time scale given by the magnetic field strength,
\begin{equation}
t'_{p,{\rm acc}}=\frac{E_p'}{B'ec}\,.
\end{equation}

Protons continue to be accelerated until they lose energy faster than their acceleration time scale.
The energy loss mechanisms that they may suffer are listed in the following equations.
Protons lose energy to synchrotron losses in magnetic field,
\begin{equation}
t'_{p,{\rm sync}}=\frac{3m_p^4c^38\pi}{4\sigma_Tm_e^2E_p'B'^2}\,,
\end{equation}
where $\sigma_T=6.65\e{-25}$ cm$^2$ is the Thomson cross section.
Protons are cooled by inverse Compton scattering, which we split into two regimes,
\begin{equation}
t'_{p,{\rm IC}}=
\begin{cases}
\frac{3m_p^4c^3}{4\sigma_Tm_e^2E_p'E_\gamma'n_\gamma'}\qquad&E_p'E_\gamma'<m_p^2c^4\\
\frac{3E_p'E_\gamma'}{4\sigma_Tm_e^2c^5n_\gamma'}&E_p'E_\gamma'>m_p^2c^4
\end{cases}\,.
\end{equation}
The Bethe-Heitler process ($p\gamma\to pe^+e^-$) has a cooling time of,
\begin{equation}
t'_{p,{\rm BH}}=\frac{E_p'\sqrt{m_p^2c^4+2E_p'E_\gamma'}}{2n_\gamma'\sigma_{\rm BH}m_ec^3(E_p'+E_\gamma')}\,,
\end{equation}
where $\sigma_{\rm BH}$ is,
\begin{equation}
\sigma_{\rm BH}=\alpha r_e^2\left[\frac{28}9\ln\left(\frac{2E_p'E_\gamma'}{m_pm_ec^4}\right)-\frac{106}9\right]\,.
\label{eq:sigma BH}
\end{equation}
with $\alpha$ being the fine structure constant and $r_e$ the classical electron radius.
Protons are also cooled via $p\gamma$ and $pp$ interactions.
Their cooling times are,
\begin{align}
t'_{p,p\gamma}&=\frac{E_p'}{c\sigma_{p\gamma}n_\gamma'\Delta E_p'}\,\\
t'_{p,{\rm hc}}&=\frac{E_p'}{c\sigma_{pp}n_p'\Delta E_p'}\,,
\end{align}
with $\Delta E_p'/E_p'=0.2,0.8$ for the $p\gamma,pp$ cases respectively.
Energy dependent cross sections, $\sigma_{p\gamma}$ and $\sigma_{pp}$, are taken from \citep{Olive:2016xmw}\footnote{For low energies ($\sqrt s<10$ GeV, with $s$ being the Mandelstam variable), the PDG data are used, while the $\ln^2s$ parameterization is used for high energy interactions.}.
Finally, protons lose energy due to adiabatic cooling from the expansion of the jet, 
\begin{equation}
t'_{p,{\rm ac}}=\frac{\tilde r_j}{c\Gamma}\ .
\end{equation}

Together, the inverse of the total cooling for protons is the sum of the inverses of each individual cooling time, $t_{p,{\rm c}}^{\prime-1}=\sum_it_{p,i}^{\prime-1}$.
The cooled proton spectrum is the uncooled spectrum scaled by an additional factor of $[1-\exp\left(-\eta_p\right)]$, where the number of energy losses is, $\eta_p=t'_{p,{\rm c}}/t'_{p,{\rm acc}}$.

\begin{figure}
\centering
\includegraphics[width=0.49\textwidth]{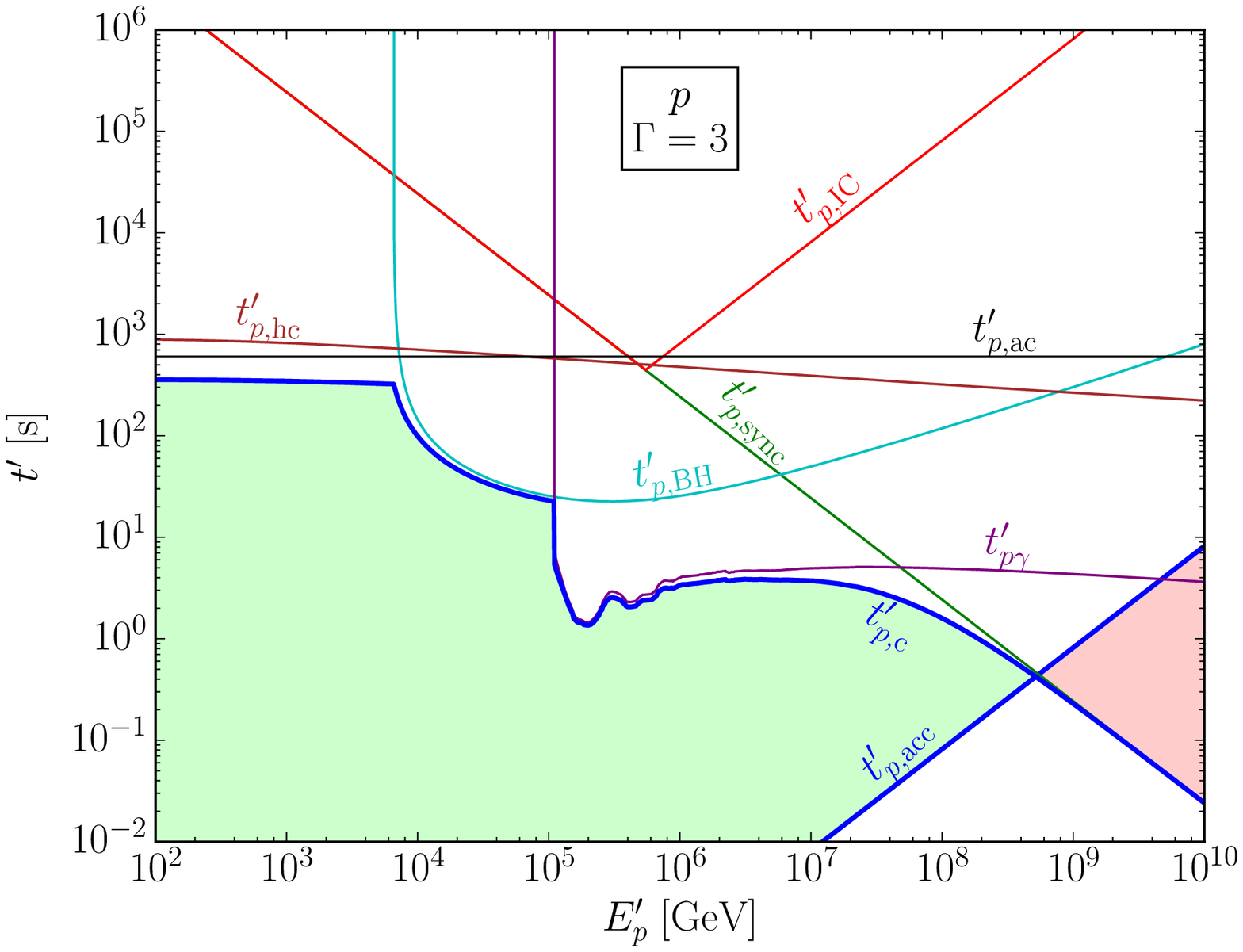}
\includegraphics[width=0.49\textwidth]{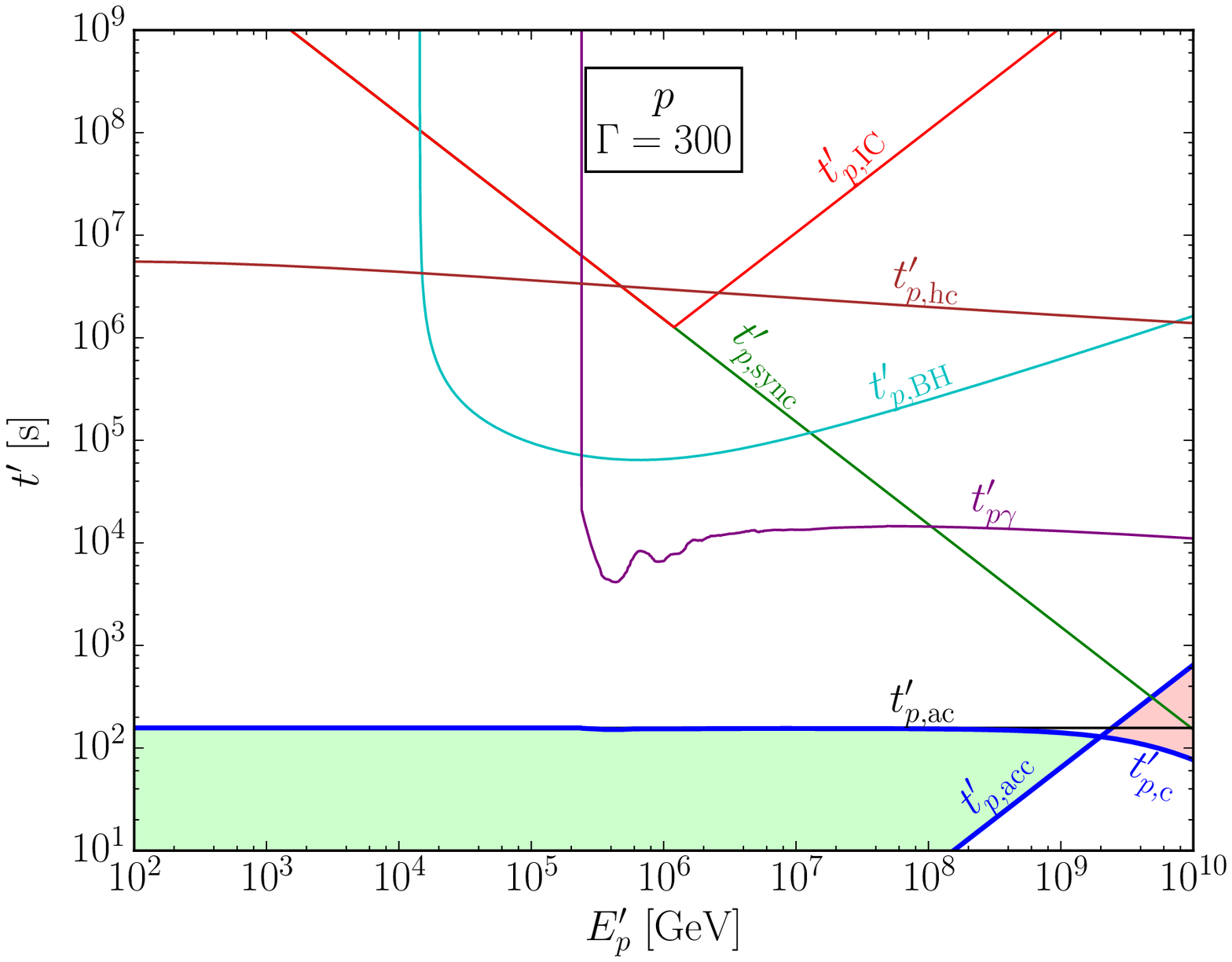}
\caption{Cooling processes for protons as a function of proton energy in the comoving frame for $\Gamma=3$ (top) and $\Gamma=300$ (bottom) for the simple GRB model.
The jet energy is fixed to $\tilde E_j=10^{51}$ erg, the energy fractions are $\eps_e=\eps_B=0.1$, and the redshift is $z=1$.
The thin solid lines mark the various individual cooling processes; the tick solids line are the total cooling, and the acceleration time.
The green shaded region on the left of each figure shows the largely uncooled portion of the spectrum, while the red region on the right shows the cooled portion of the spectrum.}
\label{fig:cooling times protons}
\end{figure}

Figure~\ref{fig:cooling times protons} shows the cooling times for protons from each process for our canonical high- and low-$\Gamma$ bursts as a function of the proton energy for jet energy $\tilde E_j=10^{51}$~erg, energy fractions $\eps_e=\eps_B=0.1$, and redshift $z=1$. The solid lines mark the various cooling processes, while the dash-dotted line represents the total cooling, and the dotted line is the acceleration time.

\subsubsection{Intermediate Particles}
\label{sssec:intermediate cooling}
Pions and kaons created in $p\gamma$ and $pp$ interactions decay into a muon neutrino and muons.
The muons subsequently decay into a muon neutrino and an electron neutrino (and an electron):
\begin{gather}
\begin{aligned}
\pi\to{}&\mu+\nu_\mu\\
K\to{}&\mu+\nu_\mu\\
&\mu\to e+\nu_\mu+\nu_e\ .
\end{aligned}
\end{gather}
It is important to include the kaon contribution. 
In fact, even though the branching ratio to produce kaons is $\sim30$ times less than that for pion, since their maximum energy is often much higher, they dominate at high energies~\citep{Ando:2005xi,Asano:2006zzb}.

Each intermediate particle also experiences cooling in a similar fashion to protons.
The total cooling time for each of these is the same as for protons after changing $m_p\to m_i$, $i=\pi,K,\mu$, and there is no contribution from the Bethe-Heitler process or $p\gamma$.
In addition, while muons do undergo hadronic cooling, the process is negligible \citep{Bulmahn:2010qna}.
The fractional energy loss for hadronic interactions for pions and kaons is $\Delta E'_{\pi(K)}/E'_{\pi(K)}=0.8$, the same as for $pp$ interactions.

The final cooled spectra for neutrinos coming from the decay of the intermediates are modified in a similar way to the proton spectrum with a factor of $\eta^{-1}=\eta_\mu^{-1}+\eta_{\pi(K)}^{-1}+\eta_p^{-1}$, where the number of energy losses due to the intermediates is $\eta_i=t'_{i,{\rm c}}m_i/E'_i\tau_i$ and $\tau_i$ is the rest frame lifetime of the particle.
The proton parameters are calculated at the proton energy that corresponds to the given neutrino energy, related by $a_i$.
The muon term is included only for neutrinos from a muon; for neutrinos directly from the mesons, no muon term is included.
The multiplicity factors for these cooling processes are given in Table \ref{tab:ratios}.

\begin{figure*}
\centering
\includegraphics[width=0.49\textwidth]{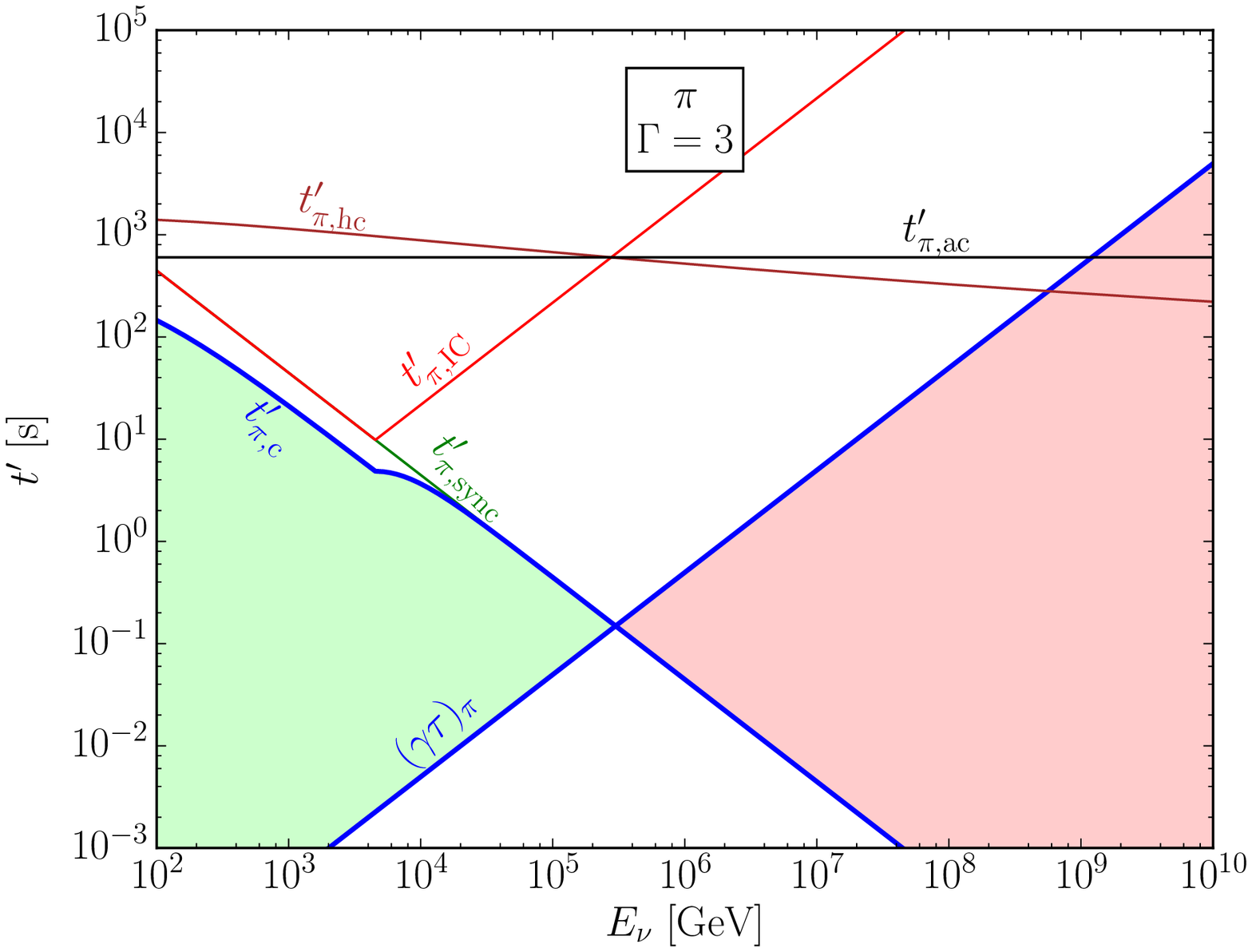}
\includegraphics[width=0.49\textwidth]{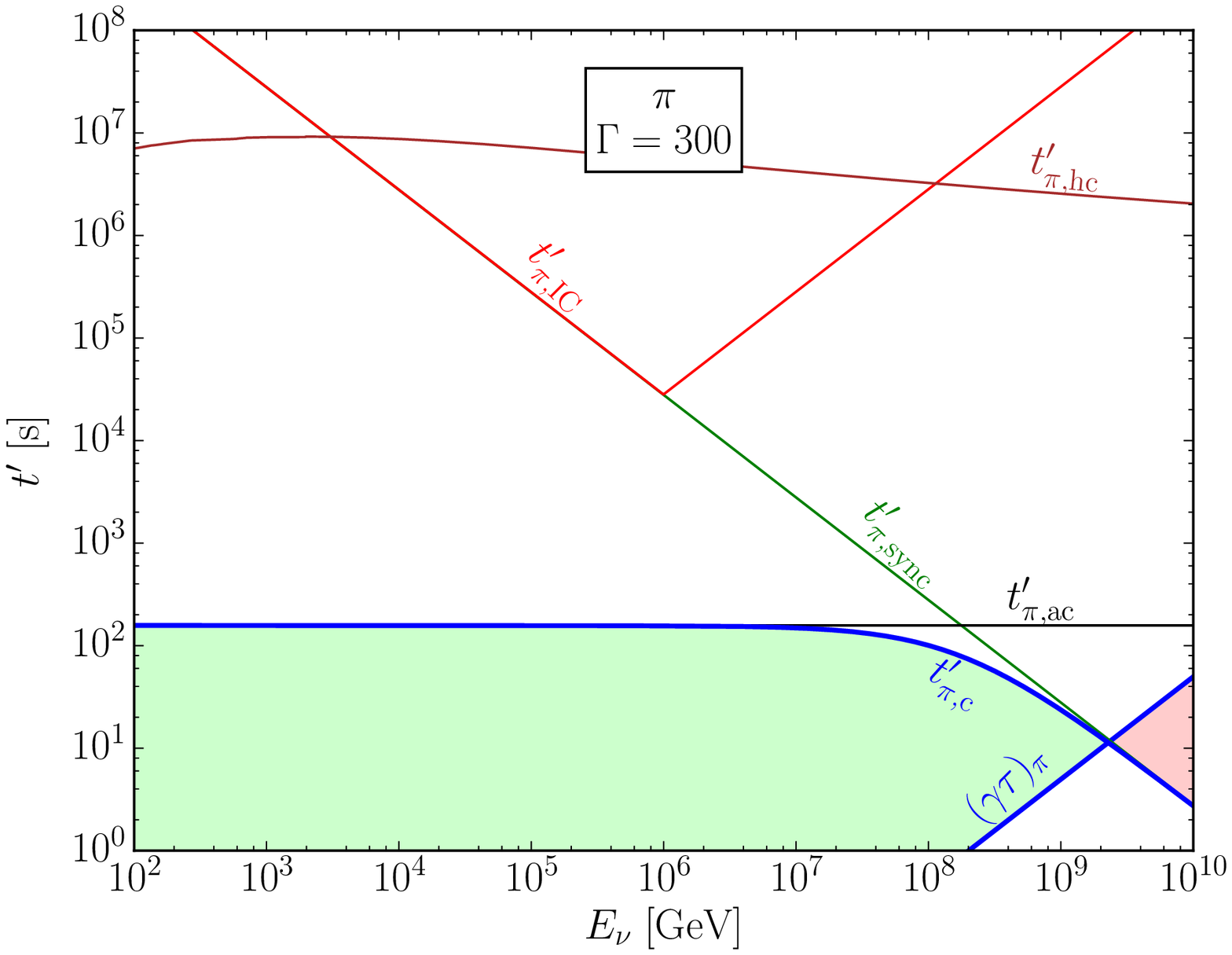}\\
\includegraphics[width=0.49\textwidth]{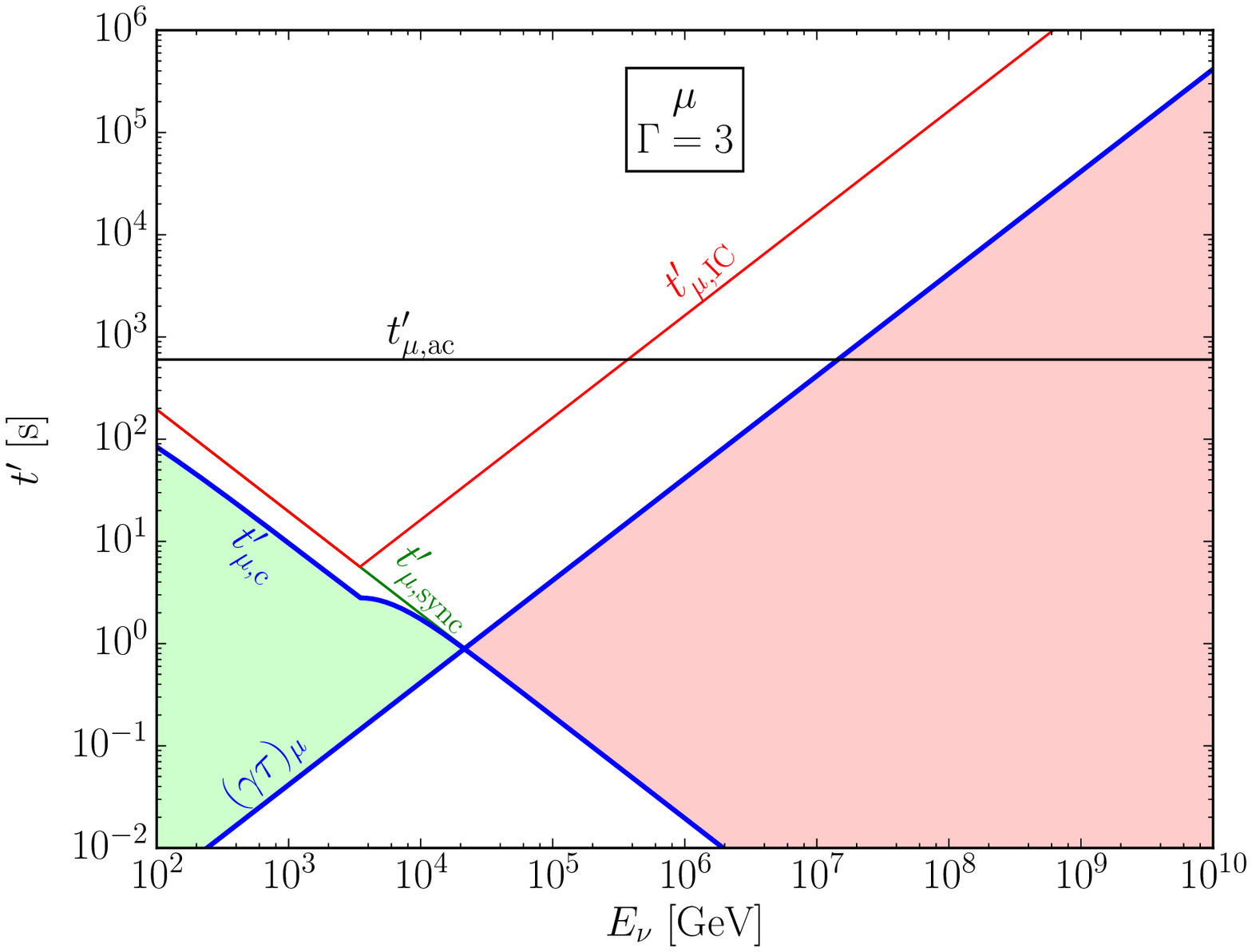}
\includegraphics[width=0.49\textwidth]{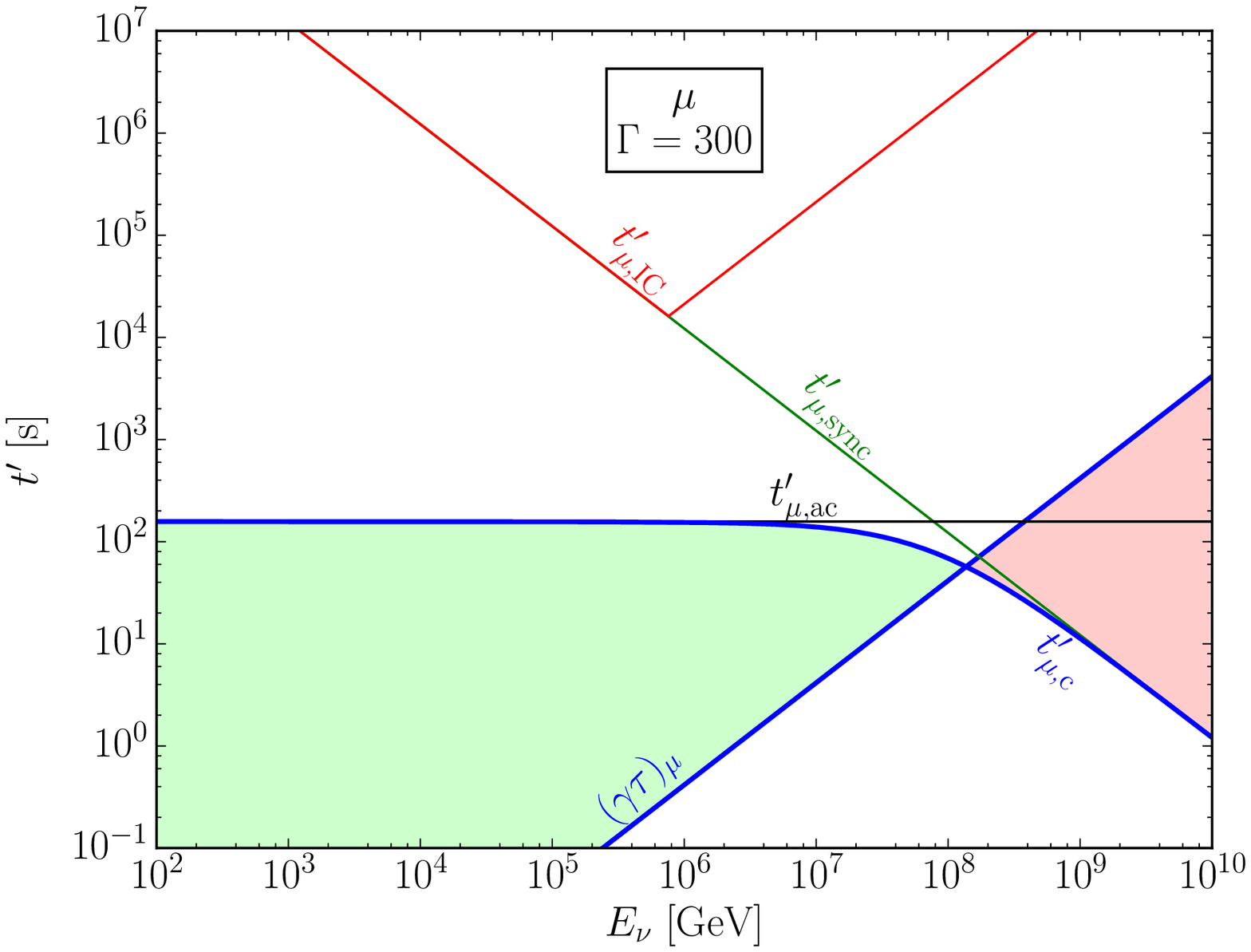}\\
\includegraphics[width=0.49\textwidth]{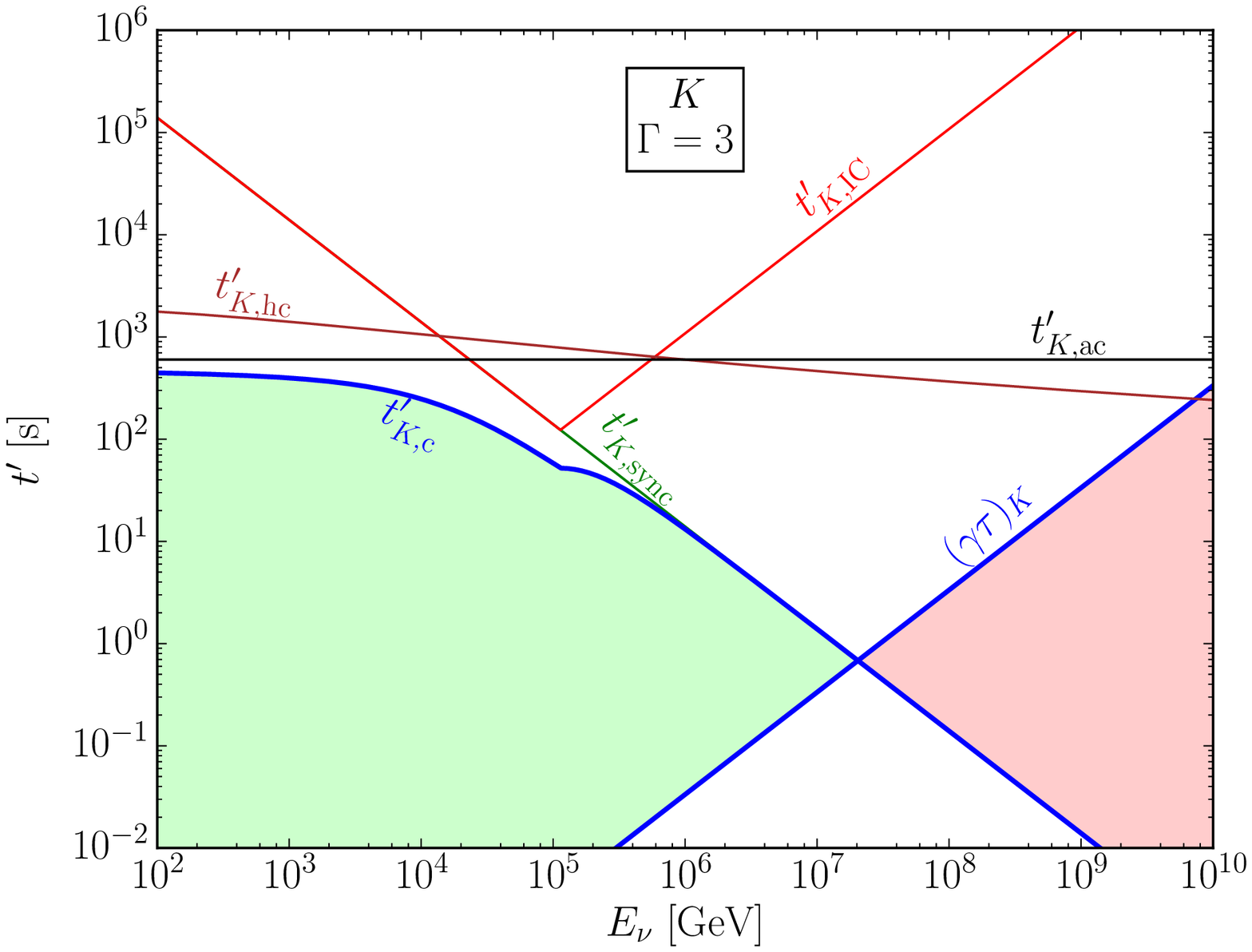}
\includegraphics[width=0.49\textwidth]{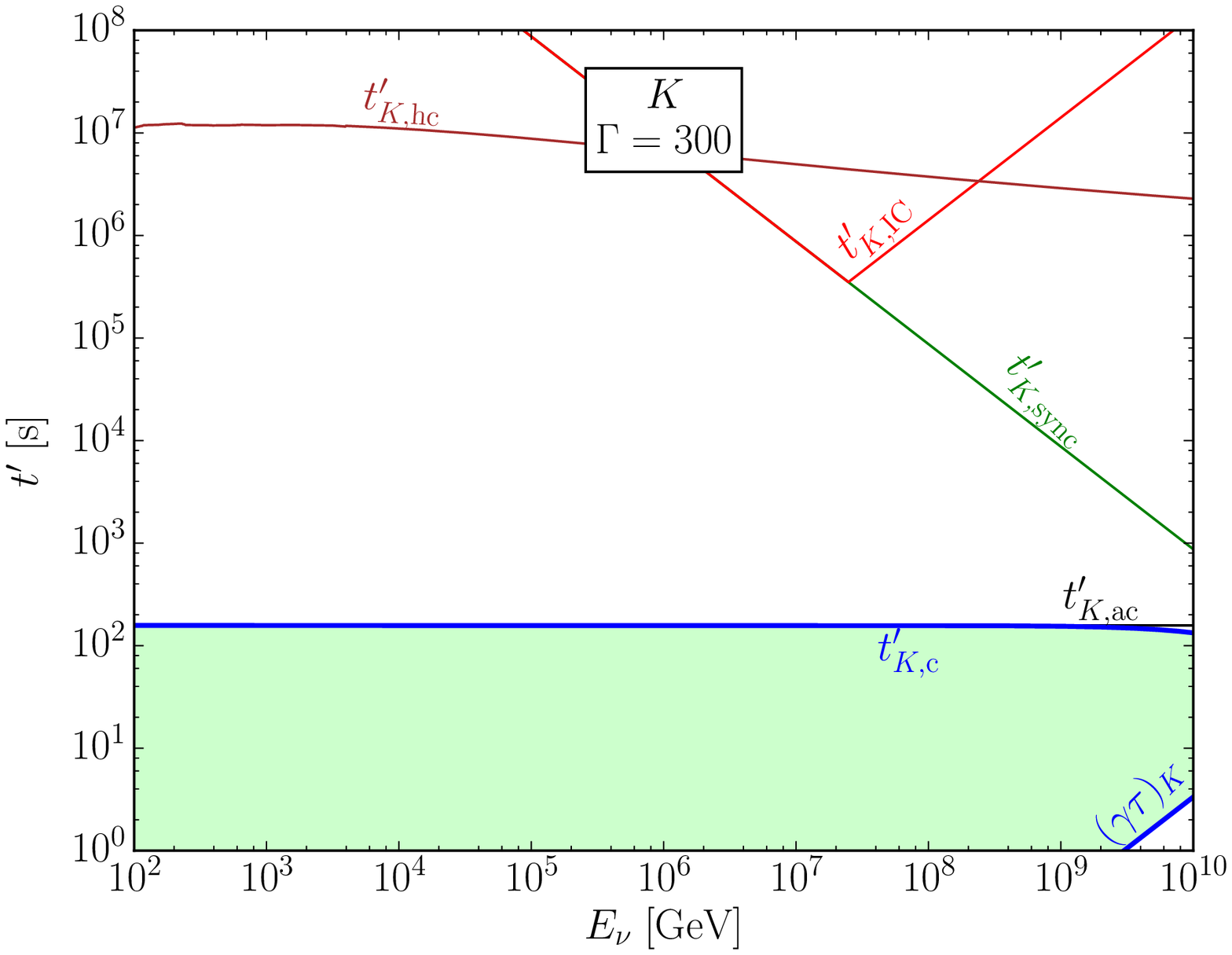}
\caption{Cooling processes for pions (top), muons (middle), and kaons (bottom) as a function of the corresponding neutrino energy in the observer frame for $\Gamma=3$ on the left and $\Gamma=300$ on the right.
The jet energy is fixed to $\tilde E_j=10^{51}$ erg, the energy fractions are $\eps_e=\eps_B=0.1$, and the redshift is $z=1$, the same as in Fig.~\ref{fig:cooling times protons}.
The thin solid lines are the various individual cooling processes; the tick lines are the total cooling, and the decay time. The green shaded region on the left of each figure shows the largely uncooled portion of the spectrum, while the red region on the right shows the cooled portion of the spectrum. 
For the adopted input parameters, synchrotron cooling is the only process that affects the spectrum.}
\label{fig:cooling times intermediates}
\end{figure*}

Figure~\ref{fig:cooling times intermediates} shows the cooling times for pions, muons, and kaons for GRB models with the same input parameters as for Fig.~\ref{fig:cooling times protons} as a function of the neutrino energy. For the adopted GRB parameters, the synchrotron cooling is the only process that affects the spectrum.

{\def\arraystretch{1.35}
\begin{table}
\centering
\caption{Ratios for different intermediate processes.
$a_i$ relates the energy of the parent proton to the energy of the resultant neutrino through a given channel, $E_{\nu,i}=a_iE_p$.
$N_i$ is the energy that goes to neutrinos from one $p\gamma$ interaction; the first number in the product is the percentage of $p\gamma$ interactions that go to $\pi,K$, and the next two numbers are the amount of energy that remain from proton to intermediate and then intermediate to neutrino.}
\begin{tabular}{c|c|c|c|c}
		&$\pi$							&$\mu_\pi$							&$K$							&$\mu_K$\\\hline
$a_i$	&$\frac15\cdot\frac14$			&$\frac15\cdot\frac34\cdot\frac13$	&$\frac15\cdot\frac12$			&$\frac15\cdot\frac12\cdot\frac13$\\
$N_i$	&$0.97\cdot\frac14\cdot\frac12$	&$0.97\cdot\frac14\cdot\frac12$		&$0.03\cdot0.63\cdot\frac12$	&$0.03\cdot0.63\cdot\frac16$
\end{tabular}
\label{tab:ratios}
\vspace*{0.06in}
\end{table}}

\subsection{Input Energy Spectra for Protons and Photons}
\label{ssec:proton and photon spectra}
We assume that the central engine is accelerating protons to high energies (where the maximum energy is determined by the acceleration time and the cooling time, see Fig.~\ref{fig:cooling times protons} and Sec.~\ref{sssec:proton cooling}).
For protons, we assume an initial Fermi shock accelerated proton spectrum, $E_p'^{-2}$.
This sets the initial power law for all subsequent spectra.

Photons are produced non-thermally inside a jet and are well described by a Band spectrum \citep{Band:1993eg},
\begin{equation}
\td{N_\gamma}{E_\gamma'}\propto
\begin{cases}
\left(\frac{E'_\gamma}{E'_{\gamma,{\rm b}}}\right)^{-1}\exp\left(-\frac{E'_\gamma}{E'_{\gamma,{\rm b}}}\right)\qquad&E'_\gamma<E'_{\gamma,{\rm b}}\\
\left(\frac{E'_\gamma}{E'_{\gamma,{\rm b}}}\right)^{-2}\exp(-1)&E'_\gamma>E'_{\gamma,{\rm b}}
\end{cases}\,,
\label{eq:Band}
\end{equation}
which is divergent as $E'_\gamma\to0$.
The photon spectrum is typically normalized over an experimentally motivated energy range, usually $E'_\gamma\in[1$ keV$,10$ MeV$]$ applied in the jet frame and so that the total photon energy in that range is $E'_{\rm iso}$. Note that in the case of optically thick sources, the photon distribution is expected to be thermal. However, in this case $\tau'_T>1$ and therefore the jet cannot accelerate particles.

The photon break energy $E'_{\gamma,{\rm b}}$ is given by the Amati relation~\citep{Amati:2002ny}:
\begin{equation}
\tilde E_{\gamma,{\rm b}}=0.364{\rm\ MeV}\left(\frac{\tilde E_{\rm iso}}{7.9\e{52}{\rm\ erg}}\right)^{0.51}\,.
\label{eq:break energy}
\end{equation}
Moreover, the Yonetoku relation relates the isotropic energy to the isotropic luminosity \citep{Yonetoku:2003gi},
\begin{equation}
\log_{10}\left(\frac{\tilde E_{\rm iso}}{10^{52}{\rm\ erg}}\right)=1.07\log_{10}\left(\frac{\tilde L_{\rm iso}}{10^{52}{\rm\ erg/s}}\right)+0.66\,.
\end{equation}
The isotropic energy is related to the jet energy by $\tilde L_{\rm iso}=(4\pi\tilde E_j\eps_e)/(0.3\tilde t_j\Omega_j)$, where the factor 0.3 is used as typically 0.3 times the peak luminosity represents the luminosity averaged over the burst duration~\citep{Kakuwa:2011aq,Liu:2012pf}.
Although the above relations have been empirically derived and are based on observed GRBs, we will assume that these relations describe all jets.

\subsection{Neutrino Energy Spectrum}
For the neutrino spectrum, we take the proton spectrum $\propto E_p^{-2}$ and multiply it by $\tau_{p\gamma},\tau_{pp}$ for $p\gamma$, $pp$ interactions respectively.
Since $\tau'_{pa}=\sigma_{pa}n'_a\tilde r_j/\Gamma$ (with $a=p,\gamma$), the effect of the photon break energy is automatically included by integrating over photon energies weighted by the photon spectrum.
Then the unnormalized uncooled (unc) neutrino spectrum from initial $pa$ interaction and intermediate $i\in\{\pi,\mu_\pi,K,\mu_K\}$ is,
\begin{equation}
\left.\td{N_\nu}{E'_\nu}\right|_{i,{\rm unc}}
\propto E^{\prime-2}_\nu\left[\int dE'_\gamma\,\td{N_\gamma}{E'_\gamma}\tau'_{p\gamma}(E'_p,E'_\gamma)+\tau'_{pp}(E'_p)\right]\,,
\end{equation}
where proton and neutrino energies are related by $a_i$ (see Table \ref{tab:ratios}) depending on which intermediate particle the neutrino comes from: $E'_p=E'_\nu/a_i$. 
The photon spectrum is normalized such that $\int dE'_\gamma\td{N_\gamma}{E'_\gamma}=1$.

\begin{figure}
\centering
\includegraphics[width=\columnwidth]{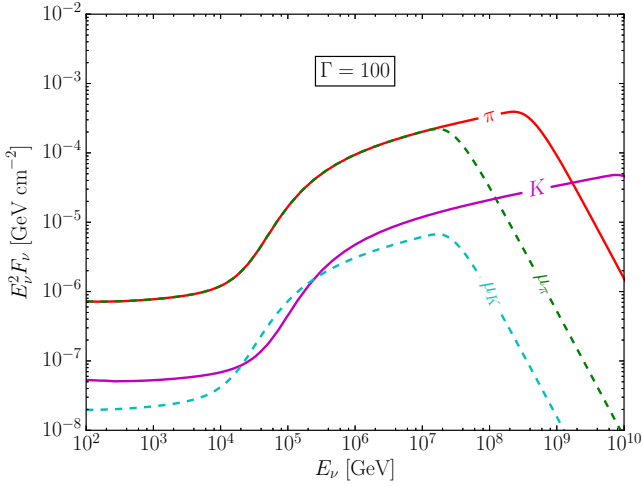}
\caption{Neutrino fluence observed at Earth for the simple GRB model as a function of the observed $E_\nu$ from one GRB with $\Gamma=100$, jet energy $\tilde E_j=10^{51}$ erg, energy fractions $\eps_e=\eps_B=0.1$, and redshift $z=1$, broken up into the different contributions coming from the intermediate particles.
The $pp$ contribution is the nearly flat part at low energies, while the $p\gamma$ component is rising at lower energies. The latter experiences the characteristic break at an energy related to the photon break energy. The cutoff energy is given by cooling and decay timescales.
The $\mu_\pi$ curve lays exactly on top of the $\pi$ curve at lower energies, a result of the fact that $a_\pi=a_{\mu_\pi}$.}
\label{fig:Fnu}
\end{figure}

We note that in the simple case where $\sigma_{p\gamma}$ is given by a step function at the $\Delta$ baryon threshold energy, we see that the $p\gamma$ correction to the $E_\nu^{\prime-2}$ part of the neutrino spectrum is $\propto E'_\nu$ before the first break and then $\propto\log E'_\nu$ after the first break until the spectrum cools to a softer spectrum after the second break.
This is different than the conventionally used correction to $E_\nu^{\prime-2}$ which is $\propto E'_\nu$, $\propto1$, $\propto E_\nu^{\prime-1}$.
Our numerical results using the full $p\gamma$ cross section confirm this behavior shown in Fig.~\ref{fig:Fnu}.

The total neutrino flux, accounting for the various energy loss mechanisms, is:
\begin{equation}
F_{\nu,i,a}(E_\nu)=\frac{(1+z)^3}{\Omega_j\Gamma d_L^2}N_af_p
\left[1-\left(1-\left.\frac{\Delta E'_p}{E'_p}\right|_a\right)^{\tau_{pa}'}\right]\left.\td{N_\nu}{E'_\nu}\right|_{i,a}\,,
\label{eq:Fnu}
\end{equation}
where $i=\pi,\mu_\pi,K,\mu_K$, $a=p,\gamma$, and the neutrino spectrum is normalized to the total jet energy, $\int dE'_\nu E'_\nu\td{N_\nu}{E'_\nu}=E'_j$.
The $[1-(1-\Delta E'_p/E'_p)^{\tau'_{pa}}]$ term accounts for the energy protons loss due to multiple $p\gamma$ and $pp$ interactions.
The remaining term accounts for energy loss from intermediate cooling, and is
\begin{equation}
f_p=\frac{\int dE'_\nu\,E'_\nu\left.\td{N_\nu}{E'_\nu}\right|_{{\rm c},pp+p\gamma}}{\int dE'_\nu\,E'_\nu\left.\td{N_\nu}{E'_\nu}\right|_{{\rm pc},pp+p\gamma}}\,,
\label{eq:fp}
\end{equation}
where pc in the denominator refers to the fact that we include only proton cooling and not the cooling of secondaries in that integral.

Figure~\ref{fig:Fnu} shows the neutrino fluence observed at Earth from one source for $\Gamma=100$, jet energy $\tilde E_j=10^{51}$~erg, energy fractions $\eps_e=\eps_B=0.1$, and redshift $z=1$ for each of the four intermediates. The contribution due to $pp$ interactions is mostly visible at low energies (the flat tail of the spectrum), while the $p\gamma$ component is rising at lower energies and then experiences the various cooling processes described above.

The per-flavor neutrino flux before flavor oscillations is,
\begin{equation}
\begin{aligned}
F_{\nu_e,{\rm unosc}}&=F_{\nu,\mu_\pi}+F_{\nu,\mu_K}\,\\
F_{\nu_\mu,{\rm unosc}}&=F_{\nu,\pi}+F_{\nu,\mu_\pi}+F_{\nu,K}+F_{\nu,\mu_K}\,.
\end{aligned}
\end{equation}
Neutrinos oscillate en route to Earth and
the distance averaged oscillated flux is~\citep{Anchordoqui:2013dnh}:
\begin{multline}
F_{\nu_\mu,{\rm osc}}=\frac14\sin^22\theta_{12}F_{\nu_e,{\rm unosc}}\\
+\frac18\left(4-\sin^22\theta_{12}\right)F_{\nu_\mu,{\rm unosc}}\,,
\end{multline}
\begin{multline}
F_{\nu_e,{\rm osc}}=\left(1-\frac12\sin^22\theta_{12}\right)F_{\nu_e,{\rm unosc}}\\
+\frac14\sin^22\theta_{12}F_{\nu_\mu,{\rm unosc}}\,,
\end{multline}
with $\theta_{12}=33.5^\circ$ \citep{Esteban:2016qun}. While neutrino oscillations through the stellar envelope do result in matter effects for $E_\nu\lesssim10$ TeV \citep{Mena:2006eq,Sahu:2010ap,Oliveros:2013apa}, in our analysis we will focus on neutrino energies larger than $\gtrsim$ 10 TeV as those are better constrained by the IceCube data; hence neutrino oscillations in the source are neglected in this work.

\begin{figure}
\centering
\includegraphics[width=\columnwidth]{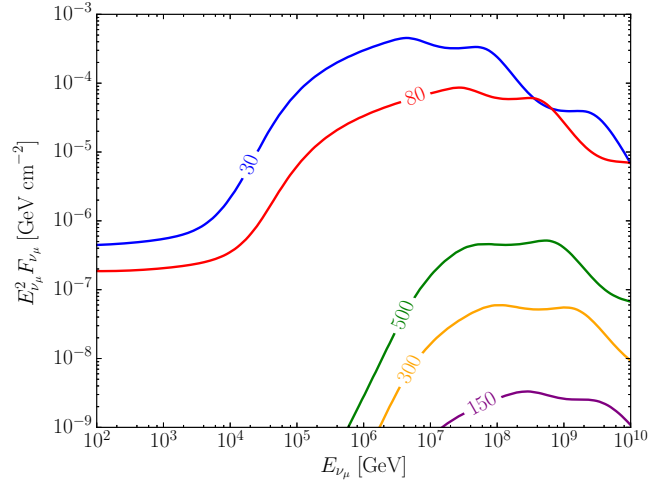}
\caption{Oscillated $\nu_\mu$ fluence for the simple GRB model as a function of the observed $E_\nu$ as seen at Earth from one GRB for different values of $\Gamma$, jet energy $\tilde E_j=10^{51}$ erg, energy fractions $\eps_e=\eps_B=0.1$, and redshift $z=1$.
The contribution from $pp$ interactions is subdominant compared to the $p\gamma$ one except at low energies.}
\label{fig:Fnu osc}
\end{figure}

Figure~\ref{fig:Fnu osc} shows the oscillated muon neutrino fluence from one source for jet energy ${\tilde E_j=10^{50}}$~erg, energy fractions $\eps_e=\eps_B=0.1$, and redshift $z=1$ as a function of the neutrino energy for different values of $\Gamma$.
As expected the $pp$ contribution, the nearly flat part at low energies, is subdominant compared to the $p\gamma$ contribution except at energies where IceCube's sensitivity to astrophysical neutrinos is low due to large atmospheric backgrounds.
Moreover, we have restricted ourselves to the high energy starting event (HESE) data set with contains neutrinos with $E_\nu\gtrsim40$ TeV and has very low backgrounds.

\section{Neutrino Diffuse Emission in the Simple GRB Model}
\label{sec:constant Gamma}
The first model we will consider is the simple GRB model where every jet has one Lorentz boost factor $\Gamma$ for the entire jet.
The population of jets will be sampled from a distribution of $\Gamma$'s that describe the data well. 

To calculate the diffuse neutrino intensity from GRBs, we assume that the GRB rate, $R(z,\Gamma)dzd\Gamma$, is separable into $R(z,\Gamma)dzd\Gamma=R(z)\xi(\Gamma)dzd\Gamma$ with,
\begin{equation}
R(z)\propto\left[(1+z)^{p_1k}+\left(\frac{1+z}{5000}\right)^{p_2k}+\left(\frac{1+z}9\right)^{p_3k}\right]^{1/k}\,,
\label{eq:SFR}
\end{equation}
where $k=-10$, $p_1=3.4$, $p_2=-0.3$, $p_3=-3.5$ are the fit parameters to the star formation rate from \citep{Yuksel:2008cu}. In fact, we assume that the CCSN rate (and in turn the rate of choked and bright GRBs) follows the star-formation rate \citep{Horiuchi:2013bc,Dahlen:2012cm}.
This function $R(z)$ is composed of three parts with power laws $p_1$, $p_2$, and $p_3$ with breaks at $z_1\approx1$ and $z_2\approx4$, and is normalized to $R(0)=1$.

We make the ansatz that $\xi(\Gamma)$ follows a power law\footnote{Note that a linear function does not in general remain non-zero when fit to the given criteria. Hence, we only use the power law parameterization.} $\xi(\Gamma)=\beta_\Gamma\Gamma^{\alpha_\Gamma}$ \citep{Tamborra:2015fzv}.
We then constrain the power law by the measured HL-GRB rate for jets with $\Gamma>200$, and the known CCSN rate for all jets:
\begin{equation}
R_{\rm SN}(0)\zeta_{\rm SN}\frac{\langle\Omega_j\rangle}{4\pi}=\int_1^{1000}d\Gamma\,\xi(\Gamma)\,,
\label{eq:RSN}
\end{equation}
\begin{equation}
\rho_{0,{\rm HL-GRB}}=\int_{200}^{1000}d\Gamma\,\xi(\Gamma)\,,
\label{eq:rhoGRB}
\end{equation}
where $R_{\rm SN}(0)\approx2\e5$ Gpc$^{-3}$ yr$^{-1}$ \citep{Dahlen:2004km,Strolger:2015kra} is the local CCSN rate, $\zeta_{\rm SN}\in(0,1]$ is the fraction of CCSNe that form jets and is taken to be redshift independent. The rate $\rho_{0,{\rm HL-GRB}}\approx0.8$ Gpc$^{-3}$ yr$^{-1}$ is an optimistic estimation for the observed local HL-GRB rate \citep{Wanderman:2009es}.
We use the range $\Gamma\in[200,1000]$ to define HL-GRB's motivated by \emph{Fermi-LAT} \citep{Bregeon:2011bu}.
The mean fraction of jets pointing toward the Earth $\langle\Omega_j\rangle/4\pi$ is
\begin{equation}
\langle\Omega_j\rangle=\frac{\int_1^{1000}d\Gamma\,\xi(\Gamma)\Omega_j}{\int_1^{1000}d\Gamma\,\xi(\Gamma)}\,,
\end{equation}
since $\Omega_j$ is a function of $\Gamma$.
For example, for our canonical GRB model with $\zeta_{\rm SN}=0.1$, we get $\alpha_\Gamma=-2.6$ and $\beta_\Gamma=6.7\e3$ Gpc$^{-3}$ yr$^{-1}$.

The resultant diffuse neutrino intensity from GRBs is,
\begin{equation}
I_\nu(E_\nu)=\int_1^{1000}d\Gamma\,\int_{z_{\mathrm{min}}}^{{z_{\mathrm{max}}}}dz
\frac{cd_L^2(1+z)^{-3}}{H_0E(z)}R(z,\Gamma)F_\nu(E_\nu)\,,
\label{eq:Inu}
\end{equation}
with $d_L(z)$ the luminosity distance, $E(z)=\sqrt{\Omega_M(1+z)^3+\Omega_\Lambda}$~\citep{Hogg:1999ad} computed by taking $\Omega_M=0.31$, $\Omega_\Lambda=0.69$, $H_0=68$ km/s/Mpc \citep{Ade:2015xua}, and $[z_{\mathrm{min}},z_{\mathrm{max}}]=[0,10]$.
We require $\theta_j<\pi/2$ in Eq.~\ref{eq:Inu}; if the jet would be larger than that we set the flux to zero.

\begin{figure}
\centering
\includegraphics[width=\columnwidth]{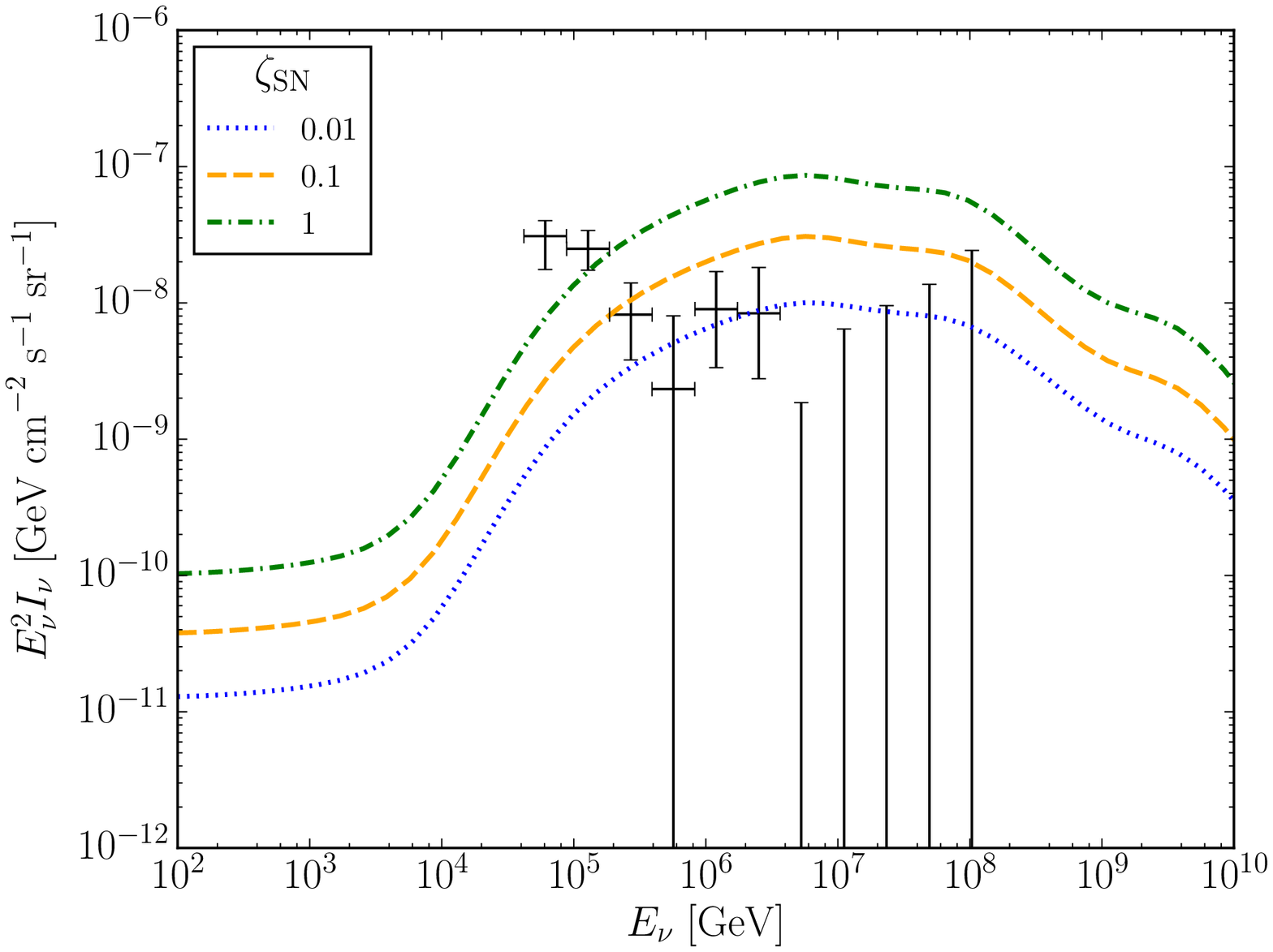}\\
\includegraphics[width=\columnwidth]{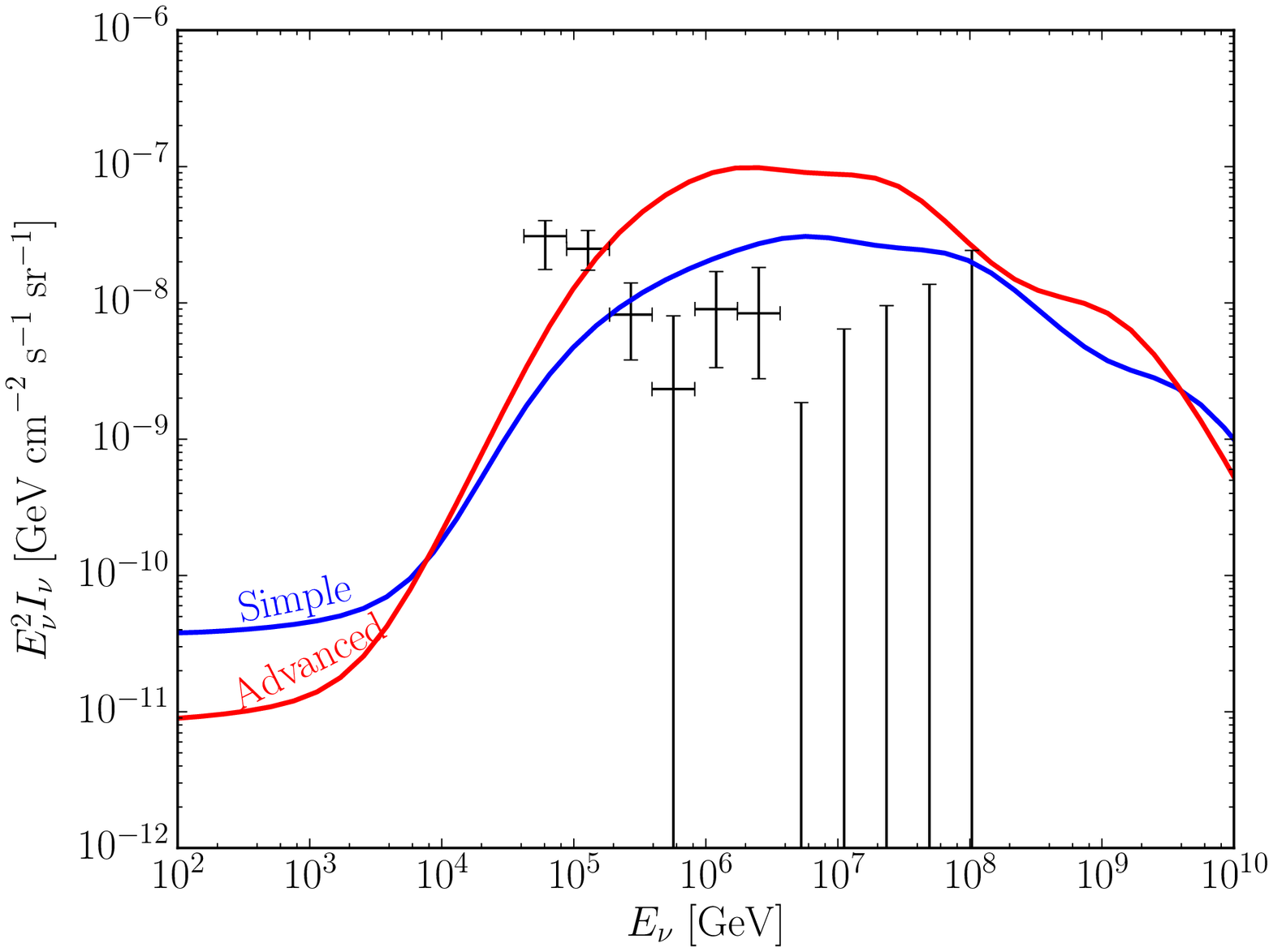}
\caption{Diffuse neutrino intensities for GRBs with jet energy ${\tilde E_j=10^{51}}$~erg, energy fractions $\eps_e=\eps_B=0.1$ plotted against the six year HESE data from IceCube \citep{Aartsen:2015zva}.
Top panel: Diffuse neutrino intensity from the simple GRB model obtained for different values of $\zeta_{\rm SN}$ ($\zeta_{\rm SN} = 0.01$ in dotted blue, $0.1$ in dashed orange and $1$ for the dash-dotted green curve).
The expected diffuse intensity is larger for larger values of $\zeta_{\rm SN}$.
Bottom panel: Diffuse neutrino intensity obtained for the simple and advance GRB model with $\zeta_{\rm SN}=0.1$.
The overall diffuse intensity is comparable within both models.}
\label{fig:Inu Gamma}
\end{figure}

\begin{figure}
\centering
\includegraphics[width=\columnwidth]{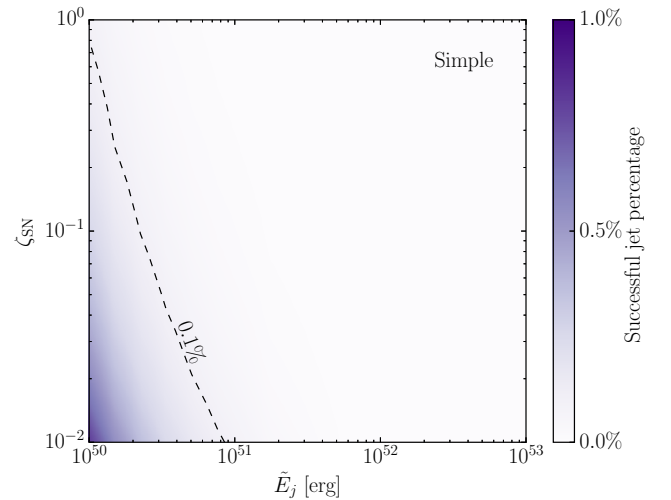}
\caption{Fraction of successful jets in the ($\tilde E_j$, $\zeta_{\rm SN}$) plane for the simple GRB model. Only a small fraction of all CCSNe can harbor a successful jet.}
\label{fig:successful}
\end{figure}

The resultant diffuse neutrino intensity is plotted in the top panel of Fig.~\ref{fig:Inu Gamma} for different values of $\zeta_{\rm SN}$ along with the six year HESE data from IceCube \citep{Aartsen:2015zva}.
Noticeably, according to $\tilde E_j$ and $\zeta_{\rm SN}$, only a small fraction of all CCSNe gives origin to a successful jet, as shown in Fig.~\ref{fig:successful}.

\section{Neutrino Diffuse Emission in the Advanced GRB Model}
\label{sec:variable Gamma}
In this Section, we take a somewhat more realistic model of the physics within a jet and allow for $\Gamma$ to vary across the jet angle from some maximum value $\Gamma_{\max}$ at the center of the jet ($\theta=0$) out to $\Gamma=1$: the advanced GRB model.
In this model, we do not rely on $\theta_j=1/\Gamma$ anymore.
Since we anticipate that multiple shocks will be accelerating the protons, the resulting $\Gamma$ distribution is a von-Mises--Fisher distribution,
\begin{equation}
\Gamma(\theta)=\Gamma_{\max}\exp\left[\kappa(\cos\theta-1)\right]\,,
\end{equation}
where the concentration $\kappa\approx1/\sigma^2$ for $\sigma$ small, with $\sigma$ the usual standard deviation.
We take $\sigma=1/{\sqrt{\Gamma_{\max}}}$ motivated by random walks of the accelerated particles within the jet.
In this model we define the volume of the jet (Eq.~\ref{eq:Volume}) with $\theta_j\to\theta_{\max}$ where $\theta_{\max}$ is defined by $\Gamma(\theta_{\max})=1$, or
\begin{equation}
\theta_{\max}=\cos^{-1}\left(1-\frac{\ln\Gamma_{\max}}\kappa\right)\,.
\end{equation}
We note that for $\kappa=\Gamma_{\max}$, $\theta_{\max}$ has a maximum at $\Gamma_{\max}=e$.
For a representative jet with $\Gamma_{\max}=300$ this corresponds to $\theta_{\max}=11^\circ$.

The component of the GRB rate $R(z,\Gamma)$ introduced in Sec.~\ref{sec:constant Gamma} depending on $\Gamma$ is assumed to follow a distribution similarly defined as in the previous section but this time this is a function of $\Gamma_{\max}$: $\xi(\Gamma_{\max})=\beta_\Gamma\Gamma_{\max}^{\alpha_\Gamma}$.

The constraint in Eq.~\ref{eq:rhoGRB} to reproduce the observed HL-GRB rate becomes,
\begin{align}
\rho_{0,{\rm HL-GRB}}={}&\int_{200}^{1000}d\Gamma_{\max}\int_{\Omega(\theta<\theta_{\max})}\frac{d\Omega}{4\pi}\xi(\Gamma_{\max})\,.
\intertext{The maximum value of $\cos\theta$ is 1 and the minimum value of $\cos\theta$ (the maximum value of $\theta$) is when $\Gamma=200$, which is $\left.\cos\theta\right|_{\min}=1-\ln(\Gamma_{\max}/200)/\kappa$.}
\rho_{0,{\rm HL-GRB}}={}&\frac{\beta_\Gamma}{\alpha_\Gamma^2}\left[(200^{\alpha_\Gamma}-1000^{\alpha_\Gamma})+\alpha_\Gamma1000^{\alpha_\Gamma}\ln\frac{1000}{200}\right]\,.
\end{align}
Similarly to Eq.~\ref{eq:RSN}, for the advanced GRB model we have
\begin{gather}
\begin{aligned}
R_{\rm SN}(0)\zeta_{\rm SN}&=\int_1^{1000}d\Gamma_{\max}\,\xi(\Gamma_{\max})\\
&=\frac{\beta_\Gamma}{\alpha_\Gamma+1}(1000^{\alpha_\Gamma+1}-1)\,,
\label{eq:RSN2}
\end{aligned}
\end{gather}
where the beaming angle factor on each side has canceled.

The diffuse neutrino intensity is
\begin{multline}
I_\nu(E_\nu)=\int_1^{1000}d\Gamma_{\max}\,\int_{\cos\theta_{\max}}^1d(\cos\theta)\,\int_{z_{\rm{min}}}^{z_{\rm{max}}}dz\\
\times\frac{cd_L^2(1+z)^{-3}}{H_0E(z)}R(z,\Gamma_{\max})F_\nu(E_\nu)\,.
\end{multline}
The $\cos\theta$ integral goes over $\theta\in[0,\theta_{\max}]$, and we again require that the jet can successfully accelerate protons.

\begin{figure}
\centering
\includegraphics[width=\columnwidth]{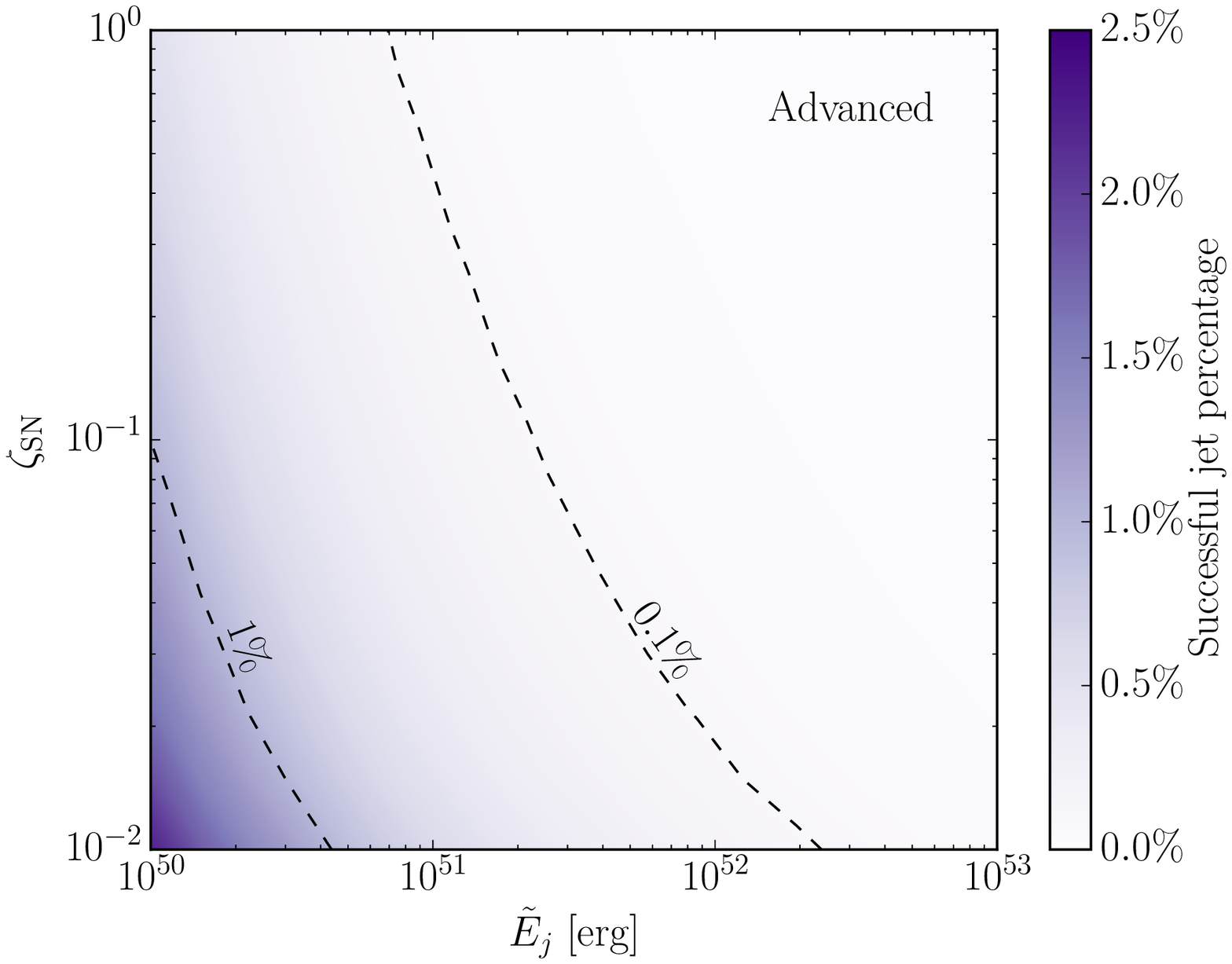}\\
\includegraphics[width=\columnwidth]{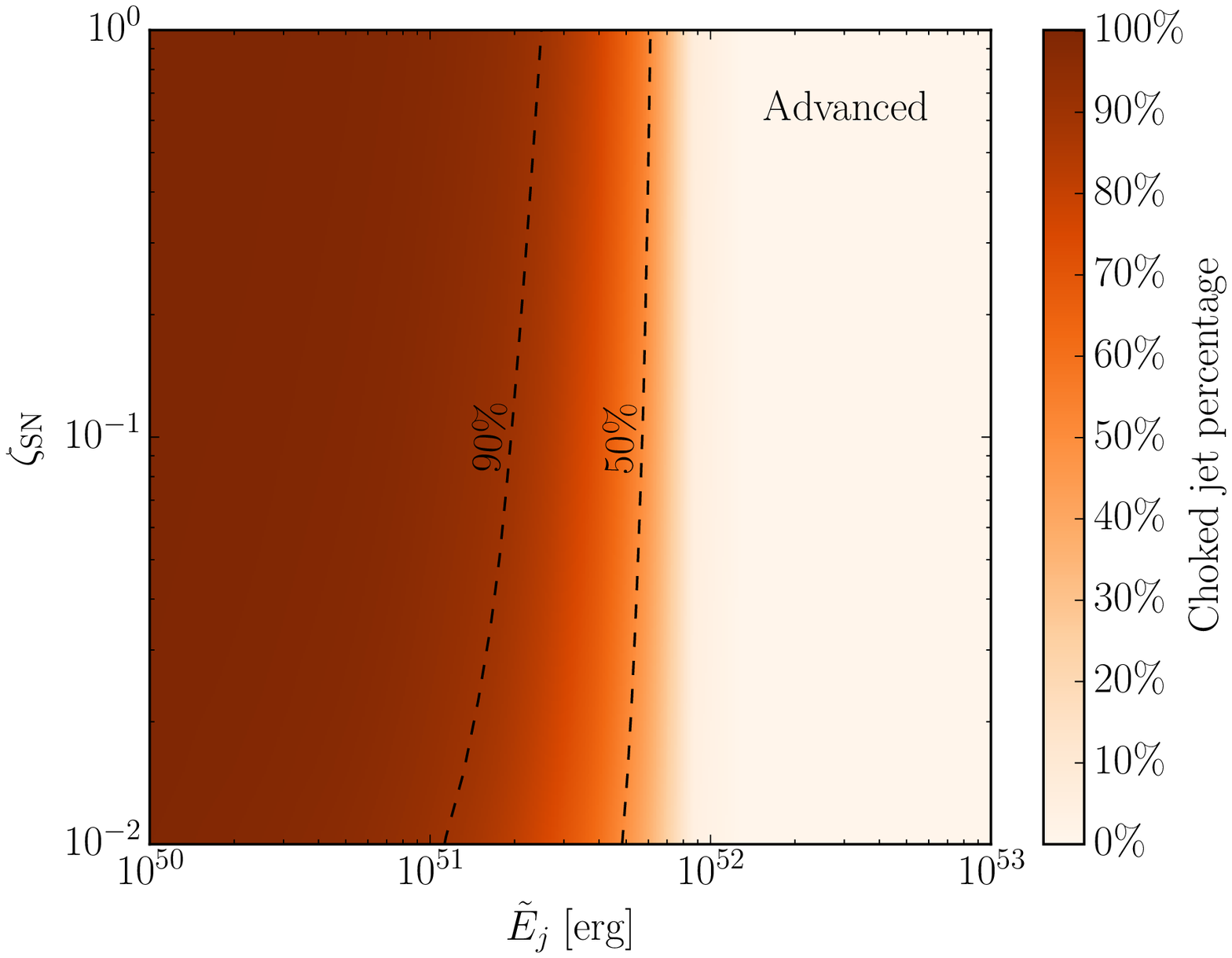}
\caption{Top panel: Fraction of successful jets in the ($\tilde E_j$, $\zeta_{\rm SN}$) plane for the advanced GRB model.
Bottom panel: Fraction of jets that are choked for the advanced GRB model.
Below $\tilde E_j\sim5\times10^{51}$ erg, 50\% of the jets are choked; below $\tilde E_j\sim2\times10^{51}$ erg, 90\% of the jets are choked.}
\label{fig:choked}
\end{figure}

The fraction of successful jets for this model is shown in the top panel of Fig.~\ref{fig:choked}. One can see that the fraction of CCSNe harboring successful jets is larger in this model with respect to the simple GRB model, given the different scaling laws intrinsic to the models.

The total diffuse neutrino intensity for our canonical GRB and $\zeta_{\rm SN}=0.1$ for all successful jets is plotted in the bottom panel of Fig.~\ref{fig:Inu Gamma} for the advanced GRB model to be compared with the one from the simple GRB model.
Both models produce comparable intensities. 

We now turn to the issue of electromagnetically choked jets. 
Depending on the jet properties, we expect that there will be some successful jets that are choked or invisible.
In the simple model, all of the jets are bright if they successfully form.
On the other hand, among all successful jets, the fraction of choked jets for the advanced model is shown in the bottom panel of Fig.~\ref{fig:choked}. Below $\tilde E_j\sim5\times10^{51}$ erg, we find that 50\% of the jets are choked; while below $\tilde E_j\sim2\times10^{51}$ erg, we obtain that 90\% of the jets are choked.
Note that, for the usually assumed typical GRB jet energy ($\tilde E_j\sim 3\times10^{51}$ erg), we expect that $\sim70\%$ of the jets are choked.

\section{IceCube Constraints on the CCSN--GRB Connection}
\label{sec:constraints}
We then construct a $\chi^2$ test with the IceCube data where we allow for neutrinos from jetted bursts to contribute a subdominant component of the observed astrophysical flux,
\begin{equation}
\chi^2=\sum_i\left[\frac{I(E_i)-I_{\rm IC}(E_i)}{I_{{\rm IC},+1\sigma}(E_i)-I_{\rm IC}(E_i)}\right]^2\Theta\left[I(E_i)-I_{\rm IC}(E_i)\right]\,,
\label{eq:chisq}
\end{equation}
where the sum is over nine energy bins in the range $[40$ TeV$,20$ PeV$]$ which includes four bins with zero events.

We do not include two of the energy bins in the $\chi^2$ that are likely under--fluctuations: the ``dip" and the ``Glashow" bins centered at $E_\nu=570$ TeV and 5.3 PeV respectively.
These bins have been investigated in the literature as possible evidence of new physics \citep{Anchordoqui:2014hua,Learned:2014vya,DiFranzo:2015qea,Tomar:2015fha}.
Under the assumption of no new physics, these bins are almost certainly under--fluctuations and unduly push up the $\chi^2$.
Moreover, IceCube has seen a through going track event with deposited energy of $2.6$ PeV~\citep{IceCube:ATEL7856}, which corresponds to a higher neutrino energy of $\sim5-10$ PeV. This would suggest that the zero bins in the data will in fact be filled in with future data.
Finally, the ``dip" bin has already begun to be filled in since the initial deficit also suggesting that it is an under--fluctuation from the first few years of data.

\begin{figure}
\centering
\includegraphics[width=\columnwidth]{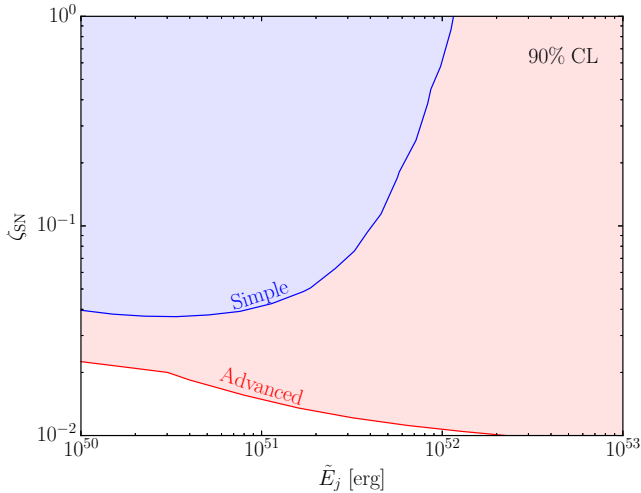}
\caption{Contour plot of the fraction of allowed bursts in the ($\tilde E_j$, $\zeta_{\rm SN}$) plane for the simple and the advanced GRB model. The shaded regions are excluded by the IceCube data \citep{Aartsen:2015rwa} at 90\% CL (7 dof). The two models lead to comparable results for $\tilde E_j\lesssim3\times10^{51}$ erg, while they differ for larger $\tilde E_j$ since most of the jets are unsuccessful at accelerating protons within the simple GRB model.
For a typical jet energy $\tilde E_j\simeq 3\times10^{51}$ erg, about $1\%$ of all CCSNe can harbor jets.}
\label{fig:Etildej_zeta}
\end{figure}

We then scan jet energies ($\tilde E_j$) and the fraction of CCSN that form jets ($\zeta_{\rm SN}$) and determine the significance of the resulting intensity given the data.
Figure~\ref{fig:Etildej_zeta} shows the 90\% contour levels for both the simple and advanced GRB models.
We note that the simple GRB model starts becoming increasingly consistent with the IceCube data above $\tilde E_j\sim10^{51}$ erg due to an increasing fraction of the jets becoming unsuccessful at accelerating protons. On the other hand, the two GRB models give comparable results for $\tilde E_j\lesssim3\times10^{51}$ erg. Our findings are discussed in the next section.

\vspace*{0.2in} 
\section{Discussion}
\label{sec:discussion}

For a typical jet energy $\tilde E_j\simeq 3\times10^{51}$ erg, Figs.~\ref{fig:choked} and \ref{fig:Etildej_zeta} suggest that less than $1\%$ of all CCSNe can harbor jets. Most of those jets are predicted to be choked.
This fraction should be compared with existing empirical constraints, based on electromagnetic observations of bright jets, suggesting that most likely a subsample of CCSNe could further evolve in jets~\citep{Guetta:2006gq,2012grb..confE.114G,Modjaz:2014doa,Margutti:2014gha,Sobacchi:2017wcq}.

It has been estimated that, under the assumption that all SN Ib/c harbor a jet, less than $10\%$ of these manages to break out from the stellar envelope and further power a prompt emission visible in $\gamma$'s~\citep{Sobacchi:2017wcq,Soderberg:2004ma,Soderberg:2005vp}. Similarly, \cite{2012grb..confE.114G} finds that the ratio of GRB to type Ib/c SNe is about $0.1-1\%$ in the local universe where type Ib/c SNe comprise up to $\sim10\%$ of all SNe.
The smaller sub-class of broadlined SNe has been linked to GRBs even in the absence of observed gamma-ray emission~\citep{Podsiadlowski:2004mt,Mazzali:2005nr,Soderberg:2005vp,Soderberg:2006tu,Milisavljevic:2014caa,Milisavljevic:2017gwq}.
Interestingly, our findings are compatible with the above observational constraints and suggest that the large majority of jets harbored in CCSNe should be not electromagnetically visible.
Additionally, \cite{Fesen:2015iqv} finds evidence that Cassiopeia A, categorized as a type IIb, may have had a weak jet, which could have either been unsuccessful at accelerating particles or choked and further suggests that all jets may be a part of a single continuous distribution.

Given the uncertainties on the CCSN explosion mechanism and concerning the conditions leading to the jet formation, we refrain from establishing a firm connection to a specific CCSN subclass in our model and instead rely on the whole CCSN population. Nevertheless, our model, being very general, does automatically take into account any eventual relativistic supernovae~\citep{Soderberg:2009ps,2006ApJ...638..930S,Chakraborti:2014dha} not directly linked to an electromagnetically bright GRB as a subclass of all stellar explosions possibly harboring jets~\citep{Margutti:2014gha,Lazzati:2011ay}.

It is worth noticing that our estimation does not take into account any dependence of the CCSN--GRB rate as a function of the redshift and progenitor metallicity~\citep{2012grb..confE.114G,Levesque:2009pt,Perley:2013fh}. 
Moreover, there may be additional metallicity/redshift dependence for the choked jets that does not apply to the electromagnetically bright GRBs. 

Our findings depend on the assumption of maximally efficient particle acceleration in the jet. The fraction of the jet energy accelerating particles and further leading to the production of neutrinos and photons is currently unconstrained; this should enter as a constant normalization factor in the estimation of the diffuse neutrino intensity. However, if particle acceleration should not be fully efficient in the jet then one should expect a correspondingly weaker upper bound on $\zeta_{\rm SN}$. The scan of different possible values of 
$\tilde E_j$ should anyway give an idea of the range of $\zeta_{\rm SN}$ compatible with the data even in the case of less efficient particle acceleration.

\section{Conclusions}
\label{sec:conclusions}
Jets harbored in core-collapse supernovae (CCSNe) are promising sources of high energy neutrinos.
By relying on the collapsar model, we assume that similar physical processes govern both electromagnetically bright and ``choked" gamma-ray bursts (GRBs).
Our calculations include neutrino production from both $p\gamma$ and $pp$ interactions and account for cooling effects of protons, pions, kaons, and muons.

Three relevant classes of jets are investigated.
These classes are based on whether or not a jet successfully accelerates protons and whether or not the jet escapes the stellar envelope.
If the jet is optically thick, then it is unsuccessful and no high energy photons or neutrinos are produced.
Successful jets can then be either visible, if photons escape the stellar envelope, or choked if the jet does not escape.
In either of these two cases, high energy neutrinos are produced. 

We calculate the neutrino diffuse intensity for two different scaling relations between the opening angle $\theta_j$ and the Lorentz boost factor $\Gamma$.
The ``simple'' GRB model assumes the classical $\theta_j=1/\Gamma$ relation, with $\Gamma$ considered to be the same throughout the jet.
On the other hand, the ``advanced'' GRB model takes a $\Gamma$ distribution throughout the jet.
$\Gamma$ is assumed to be highest along the jet axis and decrease down to one on the edges of the jet, similarly to what should occur in a more realistic case.
In the advanced GRB case, the characteristic width of the jet is given by $1/\sqrt{\Gamma_{\max}}$ where $\Gamma_{\max}$ is the Lorentz boost factor at $\theta=0$.
Our model is then tuned on the observed rate of high-luminosity GRBs and the CCSN rate. We adopt the flux of high-energy neutrinos measured by IceCube as an upper limit to the possible neutrino flux coming from bright and choked GRBs.

We find that while all of the jets in the simple GRB model are electromagnetically bright, the majority of the jets in the advanced GRB model are choked for jet energies $\tilde E_j\lesssim5\e{51}$ erg, given the differences in the scaling laws of the two GRB models.
This implies that it is crucial to adopt a refined modeling of the GRB microphysics in order to constraint the jet properties. 
In fact, for both models, the compatibility with the IceCube data \citep{Aartsen:2017wea} is similar for $\tilde E_j\lesssim10^{51}$ erg with the advanced model being slightly more constrained.
Starting above $\tilde E_j\sim10^{51}$ erg, an increasing number of jets in the simple GRB model is unsuccessful leading to a smaller diffuse intensity.

Noticeably, our findings suggest that at most $1\%$ of all CCSNe can harbor jets. Interestingly, those jets are mostly choked.
This fraction is competitive with existing empirical and observational constraints suggesting that an even smaller fraction can further lead to electromagnetically bright GRBs. 

Our study constitutes a step forward towards a realistic and general modeling of the neutrino production within bright and choked jets.
It still relies on several simplifying assumptions however, e.g.~it does not take into account any feature due to the metallicity and progenitor dependence of the CCSN population and only considers acceleration at the internal-shock radius.
As a consequence, our bounds should provide results in the correct ballpark, but may still suffer changes within a more sophisticated population-dependent modeling.

This work proves that neutrinos could be powerful messengers of the burst physics. In the light of the increasing IceCube statistics, neutrinos could provide major insights on the CCSN--GRB connection in the next future.

\acknowledgments
We are grateful to Jochen Greiner, Jens Hjorth, and Hans-Thomas Janka for useful discussions.
PBD and IT acknowledge support from the Villum Foundation (Project No.~13164), and the Danish National Research Foundation (DNRF91).
PBD thanks the Danish National Research Foundation (Grant No.~1041811001) for support during the final stages of this project. The work of IT has also been supported by the Knud H\o jgaard Foundation and the 
Deutsche Forschungsgemeinschaft through 
Sonderforschungsbereich SFB~1258 ``Neutrinos and Dark Matter
in Astro- and Particle Physics (NDM).

\bibliographystyle{apj}
\bibliography{GRB}

\begin{thebibliography}{100}
\expandafter\ifx\csname natexlab\endcsname\relax\def\natexlab#1{#1}\fi

\bibitem[{Aartsen {et~al.}(2013{\natexlab{a}})}]{Aartsen:2013jdh}
Aartsen, M.~G. {et~al.} 2013{\natexlab{a}}, Science, 342, 1242856

\bibitem[{Aartsen {et~al.}(2013{\natexlab{b}})}]{Aartsen:2013bka}
---. 2013{\natexlab{b}}, Phys. Rev. Lett., 111, 021103

\bibitem[{Aartsen {et~al.}(2014)}]{Aartsen:2014gkd}
---. 2014, Phys. Rev. Lett., 113, 101101

\bibitem[{Aartsen {et~al.}(2015{\natexlab{a}})}]{Aartsen:2015knd}
---. 2015{\natexlab{a}}, Astrophys. J., 809, 98

\bibitem[{Aartsen {et~al.}(2015{\natexlab{b}})}]{Aartsen:2014muf}
---. 2015{\natexlab{b}}, Phys. Rev., D91, 022001

\bibitem[{Aartsen {et~al.}(2015{\natexlab{c}})}]{Aartsen:2015rwa}
---. 2015{\natexlab{c}}, Phys. Rev. Lett., 115, 081102

\bibitem[{Aartsen {et~al.}(2015{\natexlab{d}})}]{Aartsen:2015zva}
Aartsen, M.~G. {et~al.} 2015{\natexlab{d}}, in {Proceedings, 34th International
  Cosmic Ray Conference (ICRC 2015): The Hague, The Netherlands, July 30-August
  6, 2015}

\bibitem[{Aartsen {et~al.}(2017{\natexlab{a}})}]{Aartsen:2017ujz}
---. 2017{\natexlab{a}}, arXiv:1707.03416 [astro-ph.HE]

\bibitem[{Aartsen {et~al.}(2017{\natexlab{b}})}]{Aartsen:2017wea}
---. 2017{\natexlab{b}}, Astrophys. J., 843, 112

\bibitem[{Ackermann {et~al.}(2011)}]{Bregeon:2011bu}
Ackermann, M. {et~al.} 2011, Astrophys. J., 729, 114

\bibitem[{Ade {et~al.}(2016)}]{Ade:2015xua}
Ade, P. A.~R. {et~al.} 2016, Astron. Astrophys., 594, A13

\bibitem[{Amati {et~al.}(2002)}]{Amati:2002ny}
Amati, L. {et~al.} 2002, Astron. Astrophys., 390, 81

\bibitem[{Anchordoqui {et~al.}(2014{\natexlab{a}})Anchordoqui, Barger,
  Goldberg, Learned, Marfatia, Pakvasa, Paul, \& Weiler}]{Anchordoqui:2014hua}
Anchordoqui, L.~A., Barger, V., Goldberg, H., Learned, J.~G., Marfatia, D.,
  Pakvasa, S., Paul, T.~C., \& Weiler, T.~J. 2014{\natexlab{a}}, Phys. Lett.,
  B739, 99

\bibitem[{Anchordoqui {et~al.}(2014{\natexlab{b}})}]{Anchordoqui:2013dnh}
Anchordoqui, L.~A. {et~al.} 2014{\natexlab{b}}, JHEAp, 1-2, 1

\bibitem[{Ando \& Beacom(2005)}]{Ando:2005xi}
Ando, S. \& Beacom, J.~F. 2005, Phys. Rev. Lett., 95, 061103

\bibitem[{Asano \& Nagataki(2006)}]{Asano:2006zzb}
Asano, K. \& Nagataki, S. 2006, Astrophys. J., 640, L9

\bibitem[{Baerwald {et~al.}(2012)Baerwald, Hummer, \& Winter}]{Baerwald:2011ee}
Baerwald, P., Hummer, S., \& Winter, W. 2012, Astropart. Phys., 35, 508

\bibitem[{Band {et~al.}(1993)}]{Band:1993eg}
Band, D. {et~al.} 1993, Astrophys. J., 413, 281

\bibitem[{Bartos {et~al.}(2013)Bartos, Beloborodov, Hurley, \&
  Márka}]{Bartos:2013hf}
Bartos, I., Beloborodov, A.~M., Hurley, K., \& Márka, S. 2013, Phys. Rev.
  Lett., 110, 241101

\bibitem[{Bromberg {et~al.}(2011{\natexlab{a}})Bromberg, Nakar, \&
  Piran}]{Bromberg:2011fm}
Bromberg, O., Nakar, E., \& Piran, T. 2011{\natexlab{a}}, Astrophys. J., 739,
  L55

\bibitem[{Bromberg {et~al.}(2011{\natexlab{b}})Bromberg, Nakar, Piran, \&
  Sari}]{Bromberg:2011fg}
Bromberg, O., Nakar, E., Piran, T., \& Sari, R. 2011{\natexlab{b}}, Astrophys.
  J., 740, 100

\bibitem[{Bulmahn(2010)}]{Bulmahn:2010qna}
Bulmahn, A.~P. 2010, PhD thesis, Iowa U.

\bibitem[{Campana {et~al.}(2006)}]{Campana:2006qe}
Campana, S. {et~al.} 2006, Nature, 442, 1008

\bibitem[{Cenko {et~al.}(2011)}]{Cenko:2010cg}
Cenko, S.~B. {et~al.} 2011, Astrophys. J., 732, 29

\bibitem[{Chakraborti {et~al.}(2015)}]{Chakraborti:2014dha}
Chakraborti, S. {et~al.} 2015, Astrophys. J., 805, 187

\bibitem[{Dahlen {et~al.}(2012)Dahlen, Strolger, Riess, Mattila, Kankare, \&
  Mobasher}]{Dahlen:2012cm}
Dahlen, T., Strolger, L.-G., Riess, A.~G., Mattila, S., Kankare, E., \&
  Mobasher, B. 2012, Astrophys. J., 757, 70

\bibitem[{Dahlen {et~al.}(2004)}]{Dahlen:2004km}
Dahlen, T. {et~al.} 2004, Astrophys. J., 613, 189

\bibitem[{Daigne \& Mochkovitch(2007)}]{Daigne:2007qz}
Daigne, F. \& Mochkovitch, R. 2007, Astron. Astrophys., 465, 1

\bibitem[{Denton {et~al.}(2017)Denton, Marfatia, \& Weiler}]{Denton:2017csz}
Denton, P.~B., Marfatia, D., \& Weiler, T.~J. 2017, JCAP, 1708, 033

\bibitem[{Dermer {et~al.}(2014)Dermer, Murase, \& Inoue}]{Dermer:2014vaa}
Dermer, C.~D., Murase, K., \& Inoue, Y. 2014, JHEAp, 3-4, 29

\bibitem[{DiFranzo \& Hooper(2015)}]{DiFranzo:2015qea}
DiFranzo, A. \& Hooper, D. 2015, Phys. Rev., D92, 095007

\bibitem[{Esteban {et~al.}(2017)Esteban, Gonzalez-Garcia, Maltoni,
  Martinez-Soler, \& Schwetz}]{Esteban:2016qun}
Esteban, I., Gonzalez-Garcia, M.~C., Maltoni, M., Martinez-Soler, I., \&
  Schwetz, T. 2017, JHEP, 01, 087

\bibitem[{Fesen \& Milisavljevic(2016)}]{Fesen:2015iqv}
Fesen, R.~A. \& Milisavljevic, D. 2016, Astrophys. J., 818, 17

\bibitem[{Gehrels \& Razzaque(2013)}]{Gehrels:2013xd}
Gehrels, N. \& Razzaque, S. 2013, Front. Phys.(Beijing), 8, 661

\bibitem[{Goldstein {et~al.}(2016)Goldstein, Connaughton, Briggs, \&
  Burns}]{Goldstein:2015fib}
Goldstein, A., Connaughton, V., Briggs, M.~S., \& Burns, E. 2016, Astrophys.
  J., 818, 18

\bibitem[{Grieco {et~al.}(2012)Grieco, Matteucci, Meynet, Longo, Della~Valle,
  \& Salvaterra}]{2012grb..confE.114G}
Grieco, V., Matteucci, F., Meynet, G., Longo, F., Della~Valle, M., \&
  Salvaterra, R. 2012, Monthly Notices of the Royal Astronomical Society, 423,
  3049

\bibitem[{Guetta \& Della~Valle(2007)}]{Guetta:2006gq}
Guetta, D. \& Della~Valle, M. 2007, Astrophys. J., 657, L73

\bibitem[{Gupta \& Zhang(2007)}]{Gupta:2006jm}
Gupta, N. \& Zhang, B. 2007, Astropart. Phys., 27, 386

\bibitem[{He {et~al.}(2012)He, Liu, Wang, Nagataki, Murase, \& Dai}]{He:2012tq}
He, H.-N., Liu, R.-Y., Wang, X.-Y., Nagataki, S., Murase, K., \& Dai, Z.-G.
  2012, Astrophys. J., 752, 29

\bibitem[{Hjorth(2013)}]{Hjorth:2013axa}
Hjorth, J. 2013, Phil. Trans. Roy. Soc. Lond., A371, 20120275

\bibitem[{Hjorth \& Bloom(2012)}]{Hjorth:2011zx}
Hjorth, J. \& Bloom, J.~S. 2012, CAPS, 51, 169

\bibitem[{Hogg(1999)}]{Hogg:1999ad}
Hogg, D.~W. 1999, arXiv:astro-ph/9905116

\bibitem[{Horiuchi \& Ando(2008)}]{Horiuchi:2007xi}
Horiuchi, S. \& Ando, S. 2008, Phys. Rev., D77, 063007

\bibitem[{Horiuchi {et~al.}(2013)Horiuchi, Beacom, Bothwell, \&
  Thompson}]{Horiuchi:2013bc}
Horiuchi, S., Beacom, J.~F., Bothwell, M.~S., \& Thompson, T.~A. 2013,
  Astrophys. J., 769, 113

\bibitem[{Kakuwa {et~al.}(2012)Kakuwa, Murase, Toma, Inoue, Yamazaki, \&
  Ioka}]{Kakuwa:2011aq}
Kakuwa, J., Murase, K., Toma, K., Inoue, S., Yamazaki, R., \& Ioka, K. 2012,
  Mon. Not. Roy. Astron. Soc., 425, 514

\bibitem[{Kumar \& Zhang(2014)}]{Kumar:2014upa}
Kumar, P. \& Zhang, B. 2014, Phys. Rept., 561, 1

\bibitem[{Lazzati {et~al.}(2012)Lazzati, Morsony, Blackwell, \&
  Begelman}]{Lazzati:2011ay}
Lazzati, D., Morsony, B.~J., Blackwell, C.~H., \& Begelman, M.~C. 2012,
  Astrophys. J., 750, 68

\bibitem[{Learned \& Weiler(2014)}]{Learned:2014vya}
Learned, J.~G. \& Weiler, T.~J. 2014, arXiv:1407.0739 [astro-ph.HE]

\bibitem[{Levesque {et~al.}(2010)}]{Levesque:2009pt}
Levesque, E.~M. {et~al.} 2010, Astrophys. J., 709, L26

\bibitem[{Liang {et~al.}(2007)Liang, Zhang, \& Dai}]{Liang:2006ci}
Liang, E., Zhang, B., \& Dai, Z.~G. 2007, Astrophys. J., 662, 1111

\bibitem[{Liu \& Wang(2013)}]{Liu:2012pf}
Liu, R.-Y. \& Wang, X.-Y. 2013, Astrophys. J., 766, 73

\bibitem[{Liu {et~al.}(2011)Liu, Wang, \& Dai}]{Liu:2011cua}
Liu, R.-Y., Wang, X.-Y., \& Dai, Z.-G. 2011, Mon. Not. Roy. Astron. Soc., 418,
  1382

\bibitem[{{L\"u} {et~al.}(2017){L\"u}, Wang, Lu, Lan, Gao, Liang, Graham,
  Zheng, Filippenko, \& Zhang}]{Lu:2017toj}
{L\"u}, H., Wang, X., Lu, R., Lan, L., Gao, H., Liang, E., Graham, M.~L.,
  Zheng, W., Filippenko, A.~V., \& Zhang, B. 2017, Astrophys. J., 843, 114

\bibitem[{MacFadyen \& Woosley(1999)}]{MacFadyen:1998vz}
MacFadyen, A. \& Woosley, S.~E. 1999, Astrophys. J., 524, 262

\bibitem[{MacFadyen {et~al.}(2001)MacFadyen, Woosley, \&
  Heger}]{MacFadyen:1999mk}
MacFadyen, A.~I., Woosley, S.~E., \& Heger, A. 2001, Astrophys. J., 550, 410

\bibitem[{Margutti {et~al.}(2014)}]{Margutti:2014gha}
Margutti, R. {et~al.} 2014, Astrophys. J., 797, 107

\bibitem[{Mazzali {et~al.}(2005)}]{Mazzali:2005nr}
Mazzali, P.~A. {et~al.} 2005, Science, 308, 1284

\bibitem[{Mena {et~al.}(2007)Mena, Mocioiu, \& Razzaque}]{Mena:2006eq}
Mena, O., Mocioiu, I., \& Razzaque, S. 2007, Phys. Rev., D75, 063003

\bibitem[{Milisavljevic \& Fesen(2017)}]{Milisavljevic:2017gwq}
Milisavljevic, D. \& Fesen, R.~A. 2017

\bibitem[{Milisavljevic {et~al.}(2015)}]{Milisavljevic:2014caa}
Milisavljevic, D. {et~al.} 2015, Astrophys. J., 799, 51

\bibitem[{Modjaz(2011)}]{Modjaz:2011bm}
Modjaz, M. 2011, Astron. J., 332, 434

\bibitem[{Modjaz {et~al.}(2014)}]{Modjaz:2014doa}
Modjaz, M. {et~al.} 2014, Astron. J., 147, 99

\bibitem[{Murase {et~al.}(2016)Murase, Guetta, \& Ahlers}]{Murase:2015xka}
Murase, K., Guetta, D., \& Ahlers, M. 2016, Phys. Rev. Lett., 116, 071101

\bibitem[{Murase \& Ioka(2013)}]{Murase:2013ffa}
Murase, K. \& Ioka, K. 2013, Phys. Rev. Lett., 111, 121102

\bibitem[{Murase {et~al.}(2006)Murase, Ioka, Nagataki, \&
  Nakamura}]{Murase:2006mm}
Murase, K., Ioka, K., Nagataki, S., \& Nakamura, T. 2006, Astrophys. J., 651,
  L5

\bibitem[{Mészáros(2006)}]{Meszaros:2006rc}
Mészáros, P. 2006, Rept. Prog. Phys., 69, 2259

\bibitem[{Mészáros(2017)}]{Meszaros:2015krr}
---. 2017, in Neutrino Astronomy- Current status, future prospects (World
  Scientific), 1--14, arXiv:1511.01396 [astro-ph.HE]

\bibitem[{Mészáros \& Waxman(2001)}]{Meszaros:2001ms}
Mészáros, P. \& Waxman, E. 2001, Phys. Rev. Lett., 87, 171102

\bibitem[{Nakar(2015)}]{Nakar:2015tma}
Nakar, E. 2015, Astrophys. J., 807, 172

\bibitem[{Osorio~Oliveros {et~al.}(2013)Osorio~Oliveros, Sahu, \&
  Sanabria}]{Oliveros:2013apa}
Osorio~Oliveros, F.~A., Sahu, S., \& Sanabria, J.~C. 2013, Eur. Phys. J., C73,
  2574

\bibitem[{Paczynski(1998)}]{Paczynski:1997yg}
Paczynski, B. 1998, Astrophys. J., 494, L45

\bibitem[{Palladino \& Vissani(2016)}]{Palladino:2016zoe}
Palladino, A. \& Vissani, F. 2016, Astrophys. J., 826, 185

\bibitem[{Patrignani {et~al.}(2016)}]{Olive:2016xmw}
Patrignani, C. {et~al.} 2016, Chin. Phys., C40, 100001

\bibitem[{Perley {et~al.}(2013)}]{Perley:2013fh}
Perley, D.~A. {et~al.} 2013, Astrophys. J., 778, 128

\bibitem[{Podsiadlowski {et~al.}(2004)Podsiadlowski, Mazzali, Nomoto, Lazzati,
  \& Cappellaro}]{Podsiadlowski:2004mt}
Podsiadlowski, P., Mazzali, P.~A., Nomoto, K., Lazzati, D., \& Cappellaro, E.
  2004, Astrophys. J., 607, L17

\bibitem[{Razzaque {et~al.}(2003{\natexlab{a}})Razzaque, Mészáros, \&
  Waxman}]{Razzaque:2002kb}
Razzaque, S., Mészáros, P., \& Waxman, E. 2003{\natexlab{a}}, Phys. Rev.
  Lett., 90, 241103

\bibitem[{Razzaque {et~al.}(2003{\natexlab{b}})Razzaque, Mészáros, \&
  Waxman}]{Razzaque:2003uv}
---. 2003{\natexlab{b}}, Phys. Rev., D68, 083001

\bibitem[{Razzaque {et~al.}(2004)Razzaque, Mészáros, \&
  Waxman}]{Razzaque:2003uw}
---. 2004, Phys. Rev., D69, 023001

\bibitem[{Sahu \& Zhang(2010)}]{Sahu:2010ap}
Sahu, S. \& Zhang, B. 2010, Res. Astron. Astrophys., 10, 943

\bibitem[{Schmid \& Turpin(2016)}]{Schmid:2015msr}
Schmid, J. \& Turpin, D. 2016, PoS, ICRC2015, 1057

\bibitem[{Schoenen \& Raedel(2015)}]{IceCube:ATEL7856}
Schoenen, S. \& Raedel, L. 2015,
  {\href{http://www.astronomerstelegram.org/?read=7856}{Astronomer's Telegram
  7856}}

\bibitem[{Senno {et~al.}(2016)Senno, Murase, \& Mészáros}]{Senno:2015tsn}
Senno, N., Murase, K., \& Mészáros, P. 2016, Phys. Rev., D93, 083003

\bibitem[{Senno {et~al.}(2017)Senno, Murase, \& Mészáros}]{Senno:2017vtd}
---. 2017, arXiv:1706.02175 [astro-ph.HE]

\bibitem[{Sobacchi {et~al.}(2017)Sobacchi, Granot, Bromberg, \&
  Sormani}]{Sobacchi:2017wcq}
Sobacchi, E., Granot, J., Bromberg, O., \& Sormani, M.~C. 2017, Mon. Not. Roy.
  Astron. Soc., 472, 616

\bibitem[{Soderberg(2006)}]{Soderberg:2006tu}
Soderberg, A.~M. 2006, AIP Conf. Proc., 836, 380, [,380(2006)]

\bibitem[{Soderberg {et~al.}(2004)Soderberg, Frail, \&
  Wieringa}]{Soderberg:2004ma}
Soderberg, A.~M., Frail, D.~A., \& Wieringa, M.~H. 2004, Astrophys. J., 607,
  L13

\bibitem[{{Soderberg} {et~al.}(2006){Soderberg}, {Nakar}, {Berger}, \&
  {Kulkarni}}]{2006ApJ...638..930S}
{Soderberg}, A.~M., {Nakar}, E., {Berger}, E., \& {Kulkarni}, S.~R. 2006, \apj,
  638, 930

\bibitem[{Soderberg {et~al.}(2006)Soderberg, Nakar, \&
  Kulkarni}]{Soderberg:2005vp}
Soderberg, A.~M., Nakar, E., \& Kulkarni, S.~R. 2006, Astrophys. J., 638, 930

\bibitem[{Soderberg {et~al.}(2010)}]{Soderberg:2009ps}
Soderberg, A.~M. {et~al.} 2010, Nature, 463, 513

\bibitem[{Sonbas {et~al.}(2015)Sonbas, Dhuga, Veres, MacLachlan, Dolek,
  Ukwatta, \& Shenoy}]{Sonbas:2014jya}
Sonbas, E., Dhuga, K.~S., Veres, P., MacLachlan, G.~A., Dolek, F., Ukwatta,
  T.~N., \& Shenoy, A. 2015, Astrophys. J., 805, 86

\bibitem[{Strolger {et~al.}(2015)Strolger, Dahlen, Rodney, Graur, Riess,
  McCully, Ravindranath, Mobasher, \& Shahady}]{Strolger:2015kra}
Strolger, L.-G., Dahlen, T., Rodney, S.~A., Graur, O., Riess, A.~G., McCully,
  C., Ravindranath, S., Mobasher, B., \& Shahady, A.~K. 2015, Astrophys. J.,
  813, 93

\bibitem[{Tamborra \& Ando(2015)}]{Tamborra:2015qza}
Tamborra, I. \& Ando, S. 2015, JCAP, 1509, 036

\bibitem[{Tamborra \& Ando(2016)}]{Tamborra:2015fzv}
---. 2016, Phys. Rev., D93, 053010

\bibitem[{Toma {et~al.}(2007)Toma, Ioka, Sakamoto, \& Nakamura}]{Toma:2006iu}
Toma, K., Ioka, K., Sakamoto, T., \& Nakamura, T. 2007, Astrophys. J., 659,
  1420

\bibitem[{Tomar {et~al.}(2015)Tomar, Mohanty, \& Pakvasa}]{Tomar:2015fha}
Tomar, G., Mohanty, S., \& Pakvasa, S. 2015, JHEP, 11, 022

\bibitem[{Wanderman \& Piran(2010)}]{Wanderman:2009es}
Wanderman, D. \& Piran, T. 2010, Mon. Not. Roy. Astron. Soc., 406, 1944

\bibitem[{Woosley \& Bloom(2006)}]{Woosley:2006fn}
Woosley, S.~E. \& Bloom, J.~S. 2006, Ann. Rev. Astron. Astrophys., 44, 507

\bibitem[{Yonetoku {et~al.}(2004)Yonetoku, Murakami, Nakamura, Yamazaki, Inoue,
  \& Ioka}]{Yonetoku:2003gi}
Yonetoku, D., Murakami, T., Nakamura, T., Yamazaki, R., Inoue, A.~K., \& Ioka,
  K. 2004, Astrophys. J., 609, 935

\bibitem[{Yuksel {et~al.}(2008)Yuksel, Kistler, Beacom, \&
  Hopkins}]{Yuksel:2008cu}
Yuksel, H., Kistler, M.~D., Beacom, J.~F., \& Hopkins, A.~M. 2008, Astrophys.
  J., 683, L5

\bibitem[{Zhang {et~al.}(2012)Zhang, Fan, Shen, Xu, Zhang, Wei, Burrows, \&
  Zhang}]{Zhang:2012jc}
Zhang, B.-B., Fan, Y.-Z., Shen, R.-F., Xu, D., Zhang, F.-W., Wei, D.-M.,
  Burrows, D.~N., \& Zhang, B. 2012, Astrophys. J., 756, 190

\end{thebibliography}

\end{document}